\documentclass[twocolumn,pre,superscriptaddress,amsmath,amssymb,aps]{revtex4-1}
\usepackage{amsmath}
\usepackage{bm}
\usepackage{braket}
\usepackage{graphicx,subfigure}
\usepackage{color}
\usepackage{url}
\usepackage{hyperref}
\usepackage{tikz}

\usepackage[normalem]{ulem}

\begin{document}

\title{Jamming and percolation in the random sequential adsorption of a binary mixture on the square lattice}

\author{Sumanta Kundu}
\email{sumanta.kundu@unipd.it}
\affiliation{Dipartimento di Fisica e Astronomia ‘Galileo Galilei’, Universit\`{a} di Padova, Via Marzolo 8, I-35131, Padova, Italy}
\affiliation{INFN, Sezione di Padova, via Marzolo 8, I-35131 Padova, Italy}

\author{Henrique C. Prates}
\affiliation{Centro de F\'{i}sica Te\'{o}rica e Computacional, Faculdade de Ci\^encias, Universidade de Lisboa, 1749-016 Lisboa, Portugal} 
\affiliation{Departamento de F\'{\i}sica, Faculdade de Ci\^{e}ncias, Universidade de Lisboa, 1749-016 Lisboa, Portugal}

\author{Nuno A.~M.~Ara\'ujo}
\email{nmaraujo@fc.ul.pt}
\affiliation{Centro de F\'{i}sica Te\'{o}rica e Computacional, Faculdade de Ci\^encias, Universidade de Lisboa, 1749-016 Lisboa, Portugal} 
\affiliation{Departamento de F\'{\i}sica, Faculdade de Ci\^{e}ncias, Universidade de Lisboa, 1749-016 Lisboa, Portugal} 

\begin{abstract}
We study the competitive irreversible adsorption of a binary mixture of monomers and square-shaped particles of linear size $R$ on the square lattice. With the random sequential adsorption model, we investigate how the jamming coverage and percolation properties depend on the size ratio $R$ and relative flux $F$. We find that the onset of percolation of monomers is always lower for the binary mixture than in the case with only monomers ($R=1$). Moreover, for values $F$ below a critical value, the higher is the flux or size of the largest species, the lower is the value of the percolation threshold for monomers.
\end{abstract}

\maketitle

\section{Introduction}

Random sequential adsorption (RSA) is the classical model to study the irreversible
adsorption of particles on surfaces. In its simplest version,
particles are adsorbed sequentially and irreversibly at random positions
on a surface without overlapping previously adsorbed ones. The dynamics leads to a jammed 
state where no more particles can adsorb, as there is no more sufficiently large vacant space available. The 
model was first introduced by Flory to describe reactions
along long polymer chains \cite{Flory1939}. The first analytical solution for the model on a line
was obtained by R\'{e}nyi~\cite{Renyi1958}. 

Over the last decades, numerous extensions of the RSA model 
were proposed and studied both analytically and numerically \cite{Feder1980,Evans1993,Privman1994,Talbot2000,Cadilhe2007,Torquato2010,Kundu2018,Furlan2020,Kubiak2016}. Despite its simplicity, the model provides deep insight into 
experimentally observed phenomena, related to chemisorption on surfaces~\cite{King1974,Guo1994,Rodgers1997}, 
adsorption on membranes~\cite{Finegold1979}, adsorption of colloids 
\cite{Feder1980-2,Onoda1986,Joshi2016,Pinto2018} and of proteins 
\cite{Adamczyk2012}, Rydberg excitation~\cite{Krapivsky2020}, and ion 
implantation in semiconductors~\cite{Subashiev2007}. One of the variants of the model which still leads to surprising results is the one focusing on the competitive adsorption of two species (binary mixture) differing in shape and/or sizes~\cite{Talbot1989,Bartelt1991,Hassan2001,Hassan2002,Doty2002,Araujo2006,Subashiev2007,Lon2007,Dias2012,Dujak2019,Darjani2021}, which also includes adsorption in the presence of quenched defects
\cite{Cornette2006,Kondrat2006,Centres2015,Tarasevich2015,Ramirez2019-2,Palacios2020}.
For example, it has been shown recently by numerical simulations that the RSA model does not 
exhibit its universal critical features at the so-called jamming and
percolation points when such defects yield strong long-range spatial 
correlations~\cite{Kundu2021}. Also, a generalization of the model using specific 
particle-size distributions has also been 
considered, to study the effect on the structure of the jamming state of polydisperse mixtures with uniform
\cite{Tarjus1991,Lj2008,Lj2011}, Gaussian~\cite{Meakin1992,Adamczyk1997,Marques2012},
and power-law~\cite{Brilliantov1996,Vieira2011}.

In this paper, we study the RSA model for a binary mixture of
particles on a square lattice. Specifically, we consider a mixture of monomers and 
square-shaped particles all interacting through excluded volume. By varying the two
model parameters, namely the aspect ratio and the relative incoming flux, 
we investigate the jamming and percolation properties. We show that the presence of square-shaped 
particles favors percolation of monomers in such a way that a cluster of monomers percolates at a density of monomers which is 
always lower than the one for the case with only monomers. 

The paper is organized as follows. In section \ref{sec:model}, we describe the model. The jamming and percolation properties are 
investigated in detail in sections \ref{sec:jamcov} and \ref{sec:perc}, 
respectively. Finally, we draw some conclusions in section \ref{sec:conclusion}.

\section{Model}
\label{sec:model}
We consider the competitive adsorption of monomers and $R\times R$ particles 
on the $L \times L$ regular square lattice with periodic boundary conditions in both
directions. $R$ and $L$ are in unit of lattice sites.
The particle species are classified as $\mathrm{A}$ (monomers) and $\mathrm{B}$ (R$^2$-mers), respectively. 
$\mathrm{A}$ and $\mathrm{B}$ particles arrive on the lattice with fluxes $f_\mathrm{A}$ and $f_\mathrm{B}$, 
respectively, and we define $F=f_\mathrm{B}/f_\mathrm{A}$ as the relative flux. We consider the Random Sequential Adsorption (RSA) model, where particles attend adsorption uniformly at random positions and the adsorption is only successful if the new particle does not overlap with previously adsorbed ones. The process of adsorption is considered irreversible. Thus, the dynamics evolves towards a jammed state where no more particles can adsorb.

At any instant of time $t$, the surface coverage $\theta_i(t)$
by a given species $i$ is defined as the fraction of the surface area covered by particles of that species, i.e.,
\begin{equation}
\theta_i(t)=\frac{N_i(t)r_i^2}{L^2} \ \ ,
\label{eq:cov}
\end{equation}
where, $N_i$ is the number of particles of species $i$ on the surface at time $t$ and $r_i$ is their linear length. In general, $\theta_i(t)$ depends both on the aspect ratio $R$ and relative flux $F$. 
Below, these parameters are systematically varied to study the jamming and percolation 
transitions. 

\section{Jammed state properties}
\label{sec:jamcov}
The jamming coverage $\theta_{\mathrm{J},i} \equiv \theta_i(t=\infty)$
of species $i$ depends both on $F$ and $R$. For $F=0$, only A-particles adsorb and the RSA model of monomers is recovered. Thus, $\theta_{\mathrm{J},\mathrm{A}}=1$. In the opposite limit, $F=\infty$, only B-particles adsorb and the RSA of $k^2$-mers (for $k=R$) is recovered, with $\theta_{\mathrm{J},\mathrm{B}}<1$, for $R>1$. For finite $F>0$, the total jamming coverage ($\theta_{\mathrm{J},\mathrm{A}}+\theta_{\mathrm{J},\mathrm{B}}$) is always unitary, but $\theta_{\mathrm{J},\mathrm{A}}$ and $\theta_{\mathrm{J},\mathrm{B}}$ are in general different, with $\theta_{\mathrm{J},\mathrm{A}}=1-\theta_{\mathrm{J},\mathrm{B}}$. 

\begin{figure}[t]
\centering
\includegraphics[width=\linewidth]{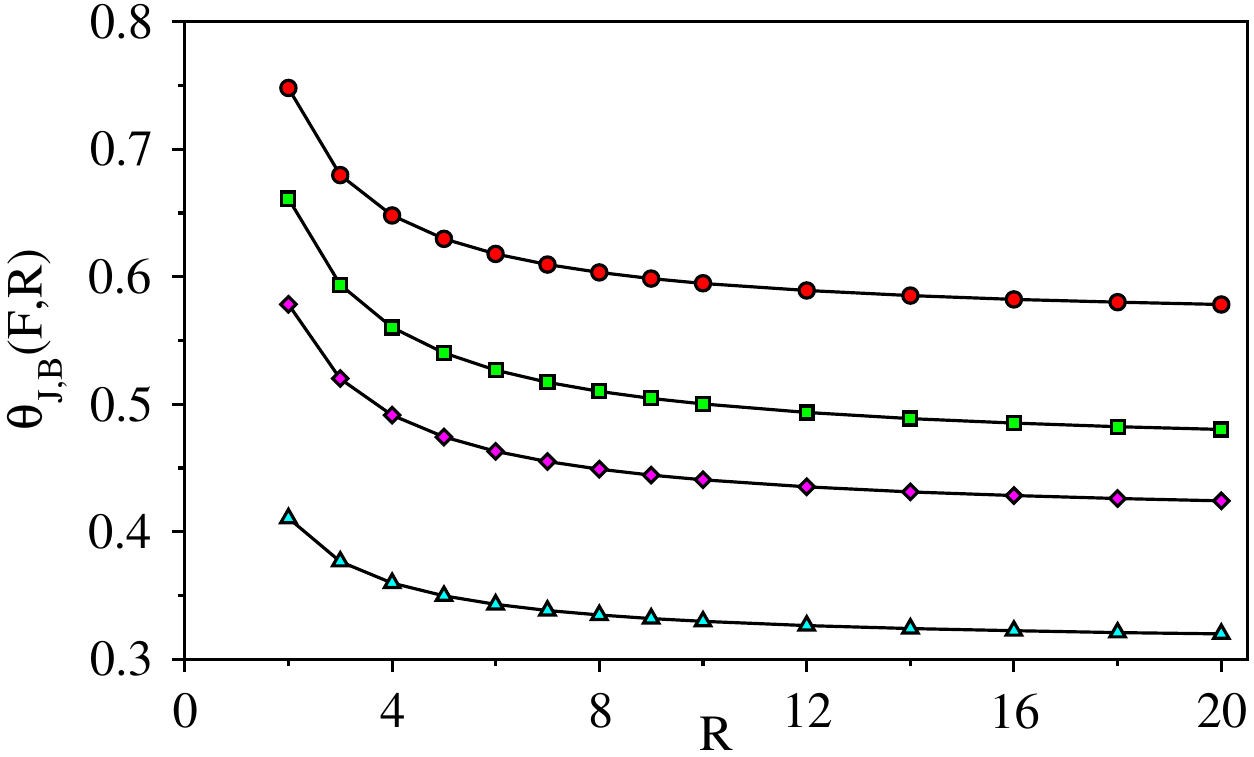}
\caption{Jamming coverage of the B-particles $\theta_{\mathrm{J},\mathrm{B}}(F,R)$ as a function of the aspect ratio $R$ for $f_\mathrm{A}=0.000$ (circles), $0.125$ 
(squares), $0.250$ (diamonds), and $0.500$ (triangles), i.e., $F=\infty$, $7$, $3$, and
$1$, respectively, on a square lattice of linear length $L=1024$. The results are averages over, at 
least $2 \times 10^5$ independent samples. The data is fitted (solid curves) with 
Eq.~\ref{eq:jamcov}, whose parameters values are $c_1=0.5621(1)$, $0.4603(1)$, $0.4077(1)$, 
and $0.3099(1)$; $c_2=0.3152(2)$, $0.3973(2)$, $0.3272(3)$, and $0.1962(2)$; and $c_3=0.1129(2)$, $0.0090(3)$, 
$0.0284(3)$, and $0.0102(3)$, respectively.}
\label{fig:cov_jam}
\end{figure}

In Fig.~\ref{fig:cov_jam} the dependence of $\theta_{\mathrm{J},\mathrm{B}}$ on $R$ is shown for four different values of $F$. The data points for each value of $F$ are consistent with the functional 
form (solid lines):
\begin{equation}
\theta_{\mathrm{J},\mathrm{B}}(F,R) = c_1 + c_2/R + c_3/R^2 \ \ ,
\label{eq:jamcov}    
\end{equation}
where $c_1$, $c_2$, and $c_3$ are fitting constants that only depend on $F$, as proposed in Ref.~\cite{Bonnier1994}. The values of the fitting constants are given in the caption of Fig.~\ref{fig:cov_jam}. It is
noteworthy that $c_1=\theta_{\mathrm{J},\mathrm{B}}(\infty,\infty)=0.5621(3)$ is equal, within statistical error bars, to the values previously reported for $k^2$-mers~\cite{Feder1980,Dickman1991,Brosilow1991,Ramirez2019}.

The monotonic decrease of $\theta_{\mathrm{J},\mathrm{B}}$ with $R$ can be explained in the following way. As the aspect ratio $R$ increases, the more sites need to be empty for a new B-particle to adsorb, while for an A-particle to adsorb is only necessary to have an empty site. Thus, the larger is the value of $R$, the more difficult it is for B-particles to adsorb, and so $\theta_{\mathrm{J},\mathrm{B}}$ decreases. Nevertheless, since the area occupied by a single B-particle also increases with $R$, for large values of $R$, the lower number of B-particles is compensated by large areas per particle and the slope of the curve for $\theta_{\mathrm{J},\mathrm{B}}$ significantly decreases.

To study the jamming transition, as in Ref.~\cite{Pasinetti2019}, we measured the standard deviation of the jamming coverage $\Delta(L)=\langle \theta_{\mathrm{J},\mathrm{B}}^2 \rangle - 
\langle \theta_{\mathrm{J},\mathrm{B}} \rangle^2$ for different system sizes, namely, $L=256$, $512$, $1024$, $2048$, and $4096$. We obtained the expected power-law decay~\cite{Pasinetti2019}, $\Delta(L) \sim L^{-1/\nu_\mathrm{J}}$,
with the critical exponent $\nu_\mathrm{J}=1$.

\section{Percolation analysis}
\label{sec:perc}
\subsection{Percolation properties of the jammed states}

In the jammed state, all sites are occupied either by A- or B-particles. In this section, we address the question how the percolation properties of the jammed configuration for
each particle species depend on the values of $F$ and $R$.  Hence, we identify and analyze the giant component of each particle species $i$, defined as the largest set of connected sites occupied by particles of that species. We consider that species $i$ percolates if the giant component spans the lattice. Note that, for a square lattice, only one species can percolate. For low values of $F$, when $f_\mathrm{A}\gg f_\mathrm{B}$, most adsorbed particles are of species A and their giant component percolates. As $F$ increases, B-particles compete with the A-particles. Beyond a critical value of $F>F_{\mathrm{c},\mathrm{B}}$ and $R<4$, it is the B-particles that percolate~\cite{Nakamura1987,Ramirez2019}. We discuss this in more
detail in the following sections.

\begin{figure}[t]
\centering
\includegraphics[width=\linewidth]{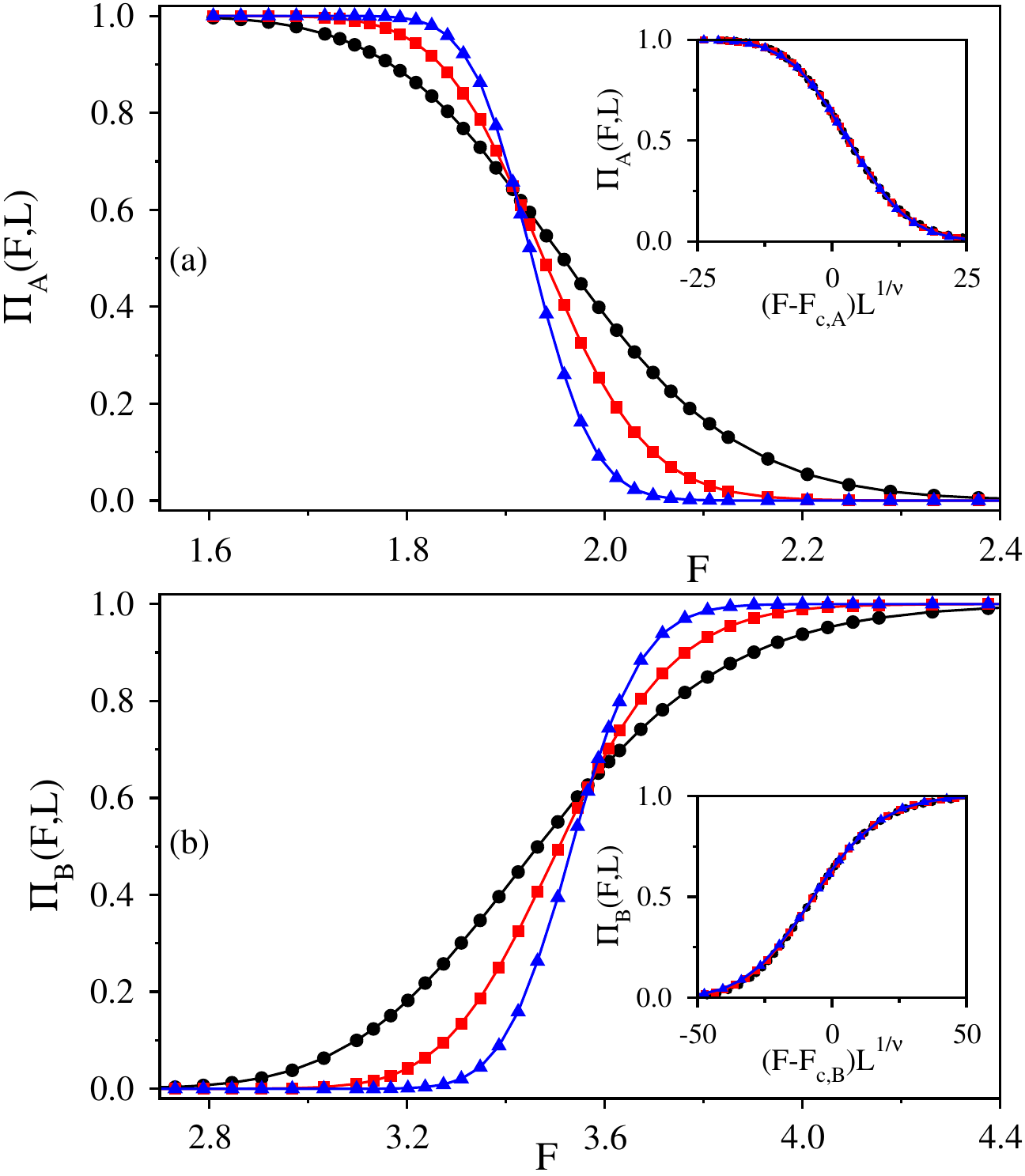}
\caption{Spanning probability for (a) species A and (b) species B in the jammed state, as a function of the relative flux $F$. The data points are averages over (at least) $10^7$, $1.8 \times 10^6,$ and $2 \times 10^5$ samples for lattices of sizes $L=256$ (circles), $512$ (squares), and $1024$ (triangles), respectively. The insets show the finite-size scaling plots for the same data.}
\label{fig:percprob}
\end{figure}

Let us start with $R=1$, which is just a competitive adsorption of two identical species. Clearly, for percolation 
to occur, the coverage of each species $i$ at the jammed state has to be, at least the percolation threshold $p_c$ for the square lattice~\cite{Jacobsen2015}, i.e., $\theta_{\mathrm{J},i}\geq p_c=0.59274605079210(2)$. One can then define a critical relative flux $F_{\mathrm{c},i}$ for each species $i$, for which $\theta_{\mathrm{J},i}=p_c$. For $R=1$, given the symmetry between A- and B-particles, $f_i=\theta_{\mathrm{J},i}$, and since $F=f_\mathrm{B}/f_\mathrm{A}$,
\begin{equation*}
\begin{cases}
F_{\mathrm{c},\mathrm{A}} = 0.68706...,\\
F_{\mathrm{c},\mathrm{B}} = 1.45547....
\end{cases}
\end{equation*}
Our numerically estimated values of $F_{\mathrm{c},\mathrm{A}}$ and $F_{\mathrm{c},\mathrm{B}}$ (details below) are in good agreement with these values.

For $R>1$, the values of $F_{\mathrm{c},\mathrm{A}}$ and $F_{\mathrm{c},\mathrm{B}}$ are estimated using numerical calculations. For each value of $F$, once the jammed state is reached, we identify the clusters of A- and B-particles separately and use the burning algorithm~\cite{Herrmann1984} to verify if any of the two span the lattice, i.e., forms a connecting path from top to bottom. Repeating this procedure 
for many independent runs, for different values of $L$ and $F$, we calculated the 
spanning probability $\Pi_\mathrm{A}(F,L)$ and $\Pi_\mathrm{B}(F,L)$ for the species A and B, respectively. 

Figures~\ref{fig:percprob}(a) and (b) show the dependence on $F$ of $\Pi_\mathrm{A}(F,L)$ and 
$\Pi_\mathrm{B}(F,L)$, for $R=2$ and $L=256$, $512$, and 
$1024$. The curves cross each other at a specific value $F=F_{\mathrm{c},i}$ for each
species $i$. A finite-size scaling analysis is performed to estimate $F_{\mathrm{c},i}$ in the thermodynamic limit (infinite lattice). In the insets of Fig.~\ref{fig:percprob} we show the data collapse for all three lattice sizes, suggesting the scaling,
\begin{equation}
\Pi_i(F,L) \sim \mathcal{G}[(F-F_{\mathrm{c},i})L^{1/\nu}]
\label{eq:scaling}
\end{equation}
for species $i$, where $\nu=0.75$ is the correlation length 
exponent of the random percolation universality class. Our best estimates for the critical
thresholds for $R=2$ are
\begin{equation*}
\begin{cases}
F_{\mathrm{c},\mathrm{A}} = 1.909(2),\\
F_{\mathrm{c},\mathrm{B}} = 3.572(2).
\end{cases}
\end{equation*}
Similarly, for $R=3$ we obtain
\begin{equation*}
\begin{cases}
F_{\mathrm{c},\mathrm{A}} = 6.346(3), \\
F_{\mathrm{c},\mathrm{B}} = 12.926(5). 
\end{cases}
\end{equation*}
\begin{figure}[t]
\centering
\includegraphics[width=\linewidth]{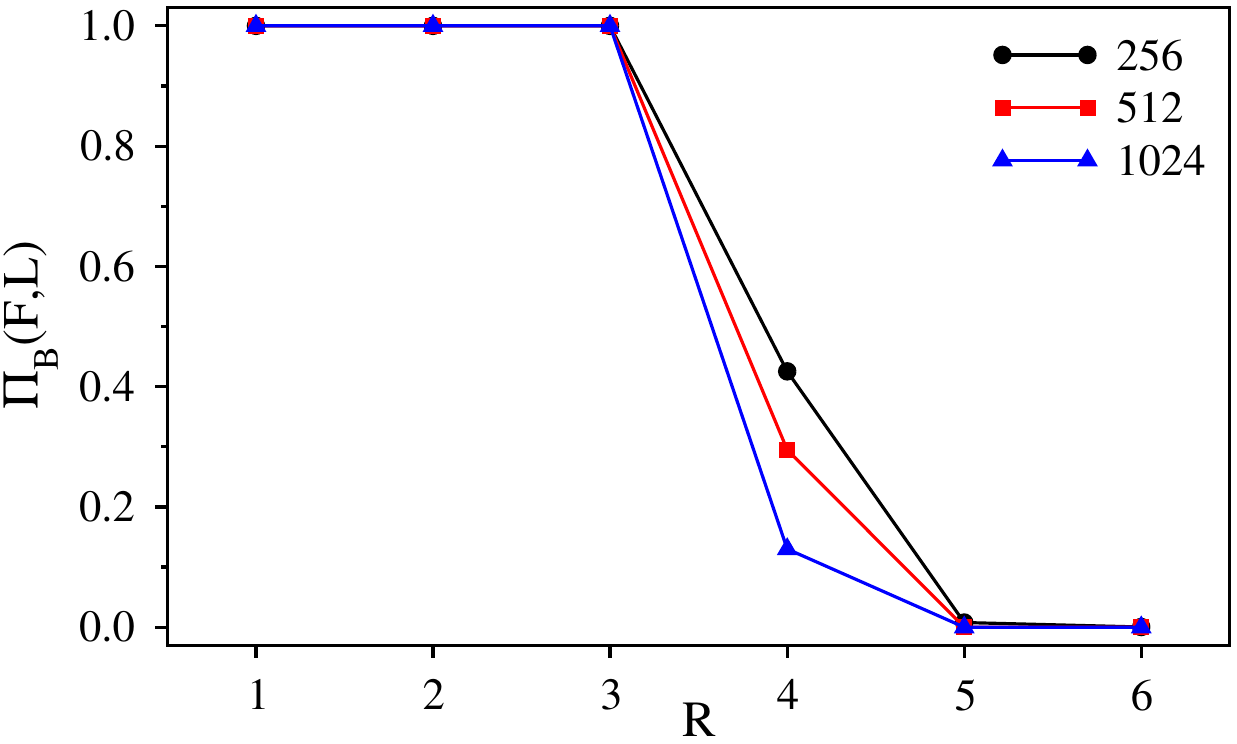}
\caption{For $F=\infty$, the spanning probability of B-particles in the 
jammed state has been plotted against the aspect ratio $R$ for $L=256$
$512$, and $1024$ using at least $8 \times 10^6, 10^6,$ and $10^5$ 
samples, respectively.}
\label{fig:Finf}
\end{figure}

For $R \ge 4$ and $F<F_{c,\mathrm{A}}$ we see that only A-particles percolate. To confirm this, we plot in Fig.~\ref{fig:Finf} the percolation probability $\Pi_\mathrm{B}(F,L)$ as a function of $R$ for $F=\infty$. It is clear that, even when there are only B-particles ($F=\infty$), percolation is only observed for $R\leq3$~\cite{Ramirez2019}. Above that value, $\Pi_\mathrm{B}(F,L)$ vanishes with $L$ and any non-zero value is just a finite-size effect. 

\begin{figure}[t]
\centering
\includegraphics[width=\linewidth]{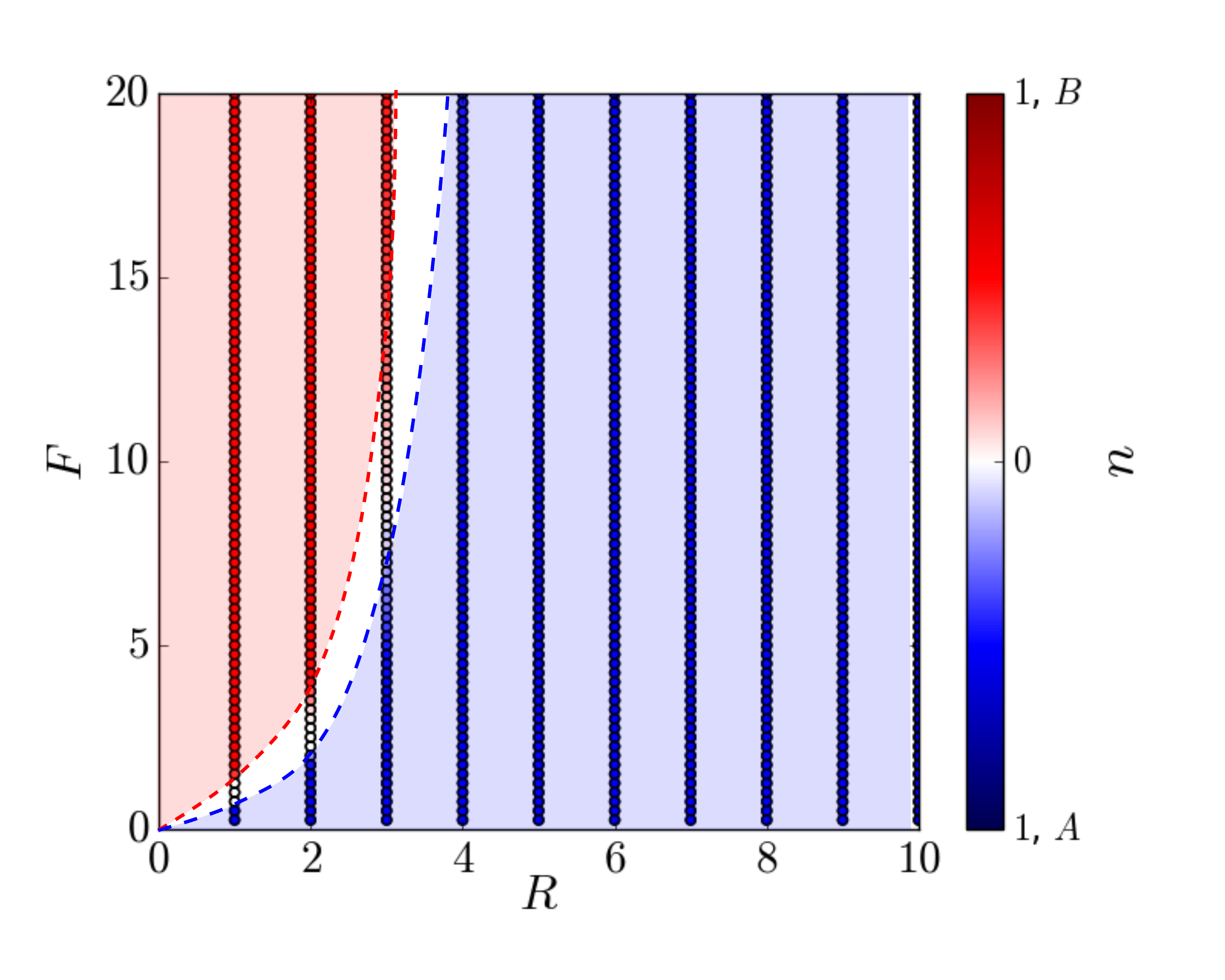}
\caption{Fraction of percolating samples for each species (A in blue, B in
red) as a function of the aspect ratio $R$ and relative flux $F$. For each
pair $(R, F)$ we performed $10^3$ independent samples on a lattice of size $L=256$.}
\label{fig:map}
\end{figure}

Performing a series of Monte Carlo simulations for different aspect ratios $R$ 
and relative fluxes $F$, we obtained a two-parameter diagram in the $F$-$R$ plane, as shown in Fig.~\ref{fig:map}. The whole
plane is divided into three separate regions. The A-species percolates 
if and only if $F<F_{\mathrm{c},\mathrm{A}}$. Similarly, B-species percolates if $F>F_{\mathrm{c},\mathrm{B}}$ and $R < 4$. In between, there is a region where 
neither the A- nor the B-species percolate. The size of this region depends on the value of $R$.

\begin{figure}[t]
\centering
\includegraphics[width=\linewidth]{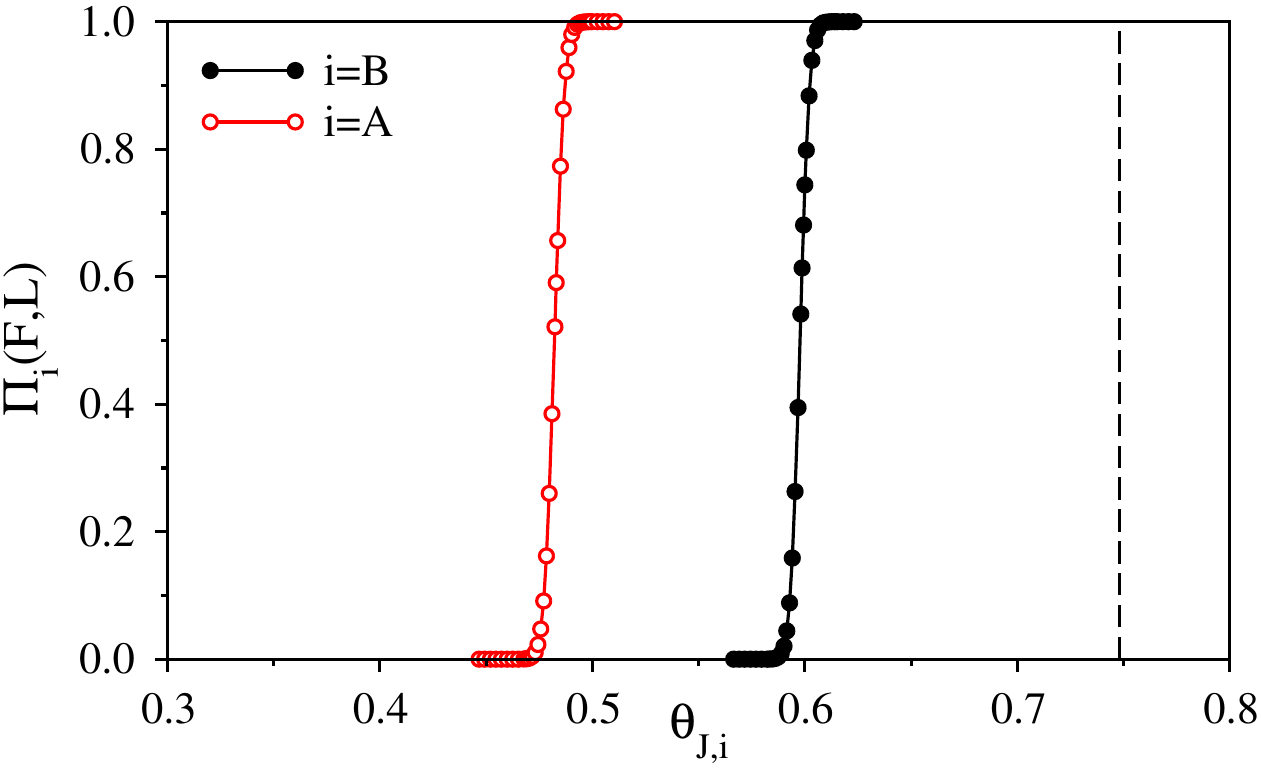}
\caption{Spanning probability $\Pi_i(F,L)$ for species $i$ as a function of the respective 
jamming coverage $\theta_{\mathrm{J},i}$ for $R=2$ and $L=1024$. The data points are based on at least $2 \times 
10^5$ independent samples. The maximum jamming coverage for the B species, i.e., 
$\theta_{\mathrm{J},\mathrm{B}}$ for $F=\infty$ is represented by the vertical dashed line.}
\label{fig:covcrit2}
\end{figure}

\begin{figure}[t]
\centering
\includegraphics[width=\linewidth]{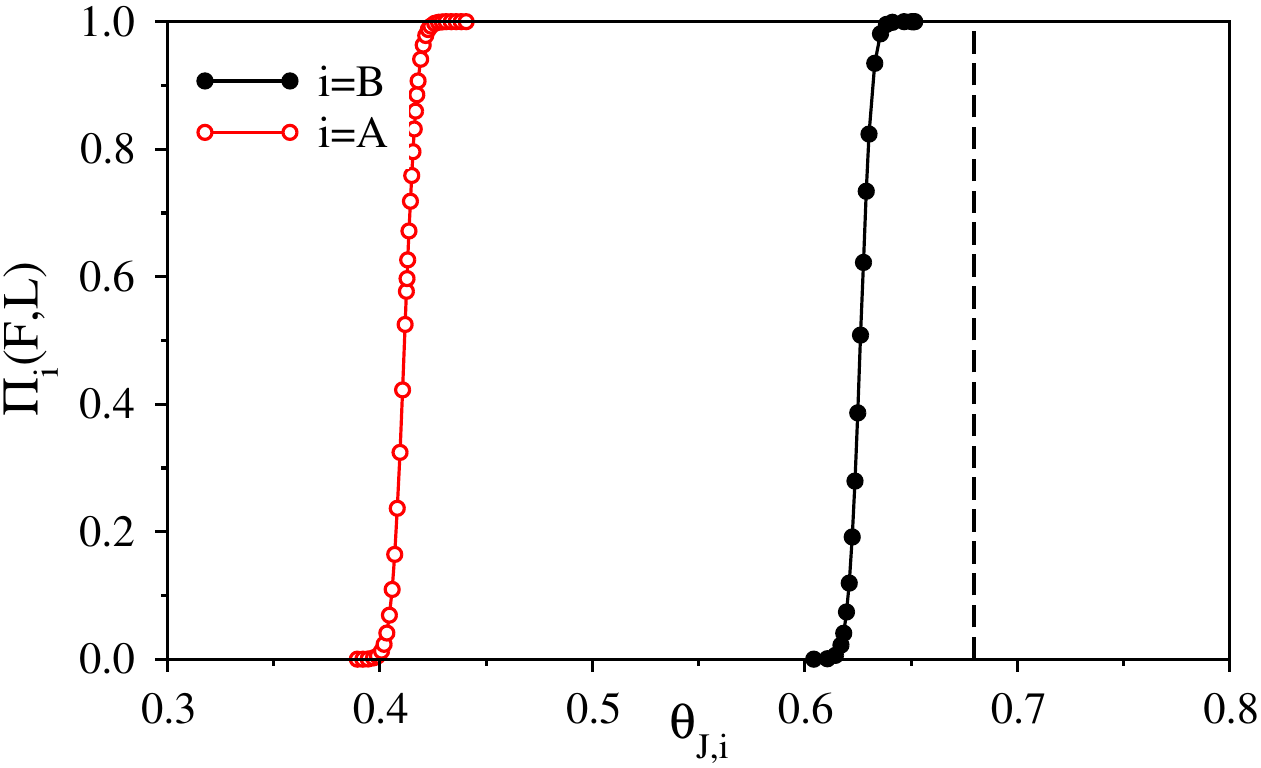}
\caption{Spanning probability $\Pi_i(F,L)$ for species $i$ as a function of the respective 
jamming coverage $\theta_{\mathrm{J},i}$ for $R=3$ and $L=1024$. The data points are based 
on at least $10^5$ independent samples. The maximum jamming coverage for the B species, 
i.e., $\theta_{\mathrm{J},\mathrm{B}}$ for $F=\infty$ is represented by the vertical dashed 
line.}
\label{fig:covcrit3}
\end{figure}

The threshold value of the jamming coverage $\theta_{\mathrm{Jc},i}$, at which the jammed configuration of each species percolates depends on the value of the flux $F$. To estimate the critical value $\theta_{\mathrm{Jc}}$, we plot $\Pi_i$ as a function of the coverage $\theta_{\mathrm{J},i}$ for each species, obtained for different values of $F$, as shown in Figs.~\ref{fig:covcrit2} and \ref{fig:covcrit3}, for $R=2$ and $R=3$, respectively, and $L=1024$. A data collapse is then obtained for three different system sizes, which is consistent with the scaling form 
of Eq.~\ref{eq:scaling} (not shown). For $R=2$, we obtain
\begin{equation*}
\begin{cases}
\theta_{\mathrm{Jc},\mathrm{A}} = 0.483(1),\\
\theta_{\mathrm{Jc},\mathrm{B}} = 0.599(1),
\end{cases}
\end{equation*}
while for $R=3$,
\begin{equation*}
\begin{cases}
\theta_{\mathrm{Jc},\mathrm{A}} = 0.413(1),\\
\theta_{\mathrm{Jc},\mathrm{B}} = 0.628(1).
\end{cases}
\end{equation*}

It is noticeable from Figs.~\ref{fig:covcrit2} and \ref{fig:covcrit3} that the curve for the spanning probability of species A shifts towards smaller value of the coverage of A as $R$ increases. One could expect 
the opposite scenario, as the presence of large B 
particles might hinder the growth of the largest component of A species, requiring more A particles for it to percolate. However, the presence of B-particles also limits the regions where A-particles can absorb. The obtained result suggests that this helps to obtain a spanning cluster of A-particles for lower values of the coverage. We discuss in the next section in more detail how the presence of large B-particles might favor percolation of monomers (A-particles).

\subsection{Percolation of monomers}\label{sec:transition}
We now focus on the percolation of monomers (A-particles). So far, we have only studied the jammed state configuration. However, in general, the percolation transition is expected to occur before jamming. Since, for $F<F_{\mathrm{c},\mathrm{A}}(R)$, the A-particles always percolate in the jammed state, for all values of $F<F_{c,A}(R)$ there exist a critical value of the coverage $\theta_{\mathrm{c},\mathrm{A}}$ for which the largest component of A-particles percolate. 
\begin{figure}[t]
\centering
\includegraphics[width=\linewidth]{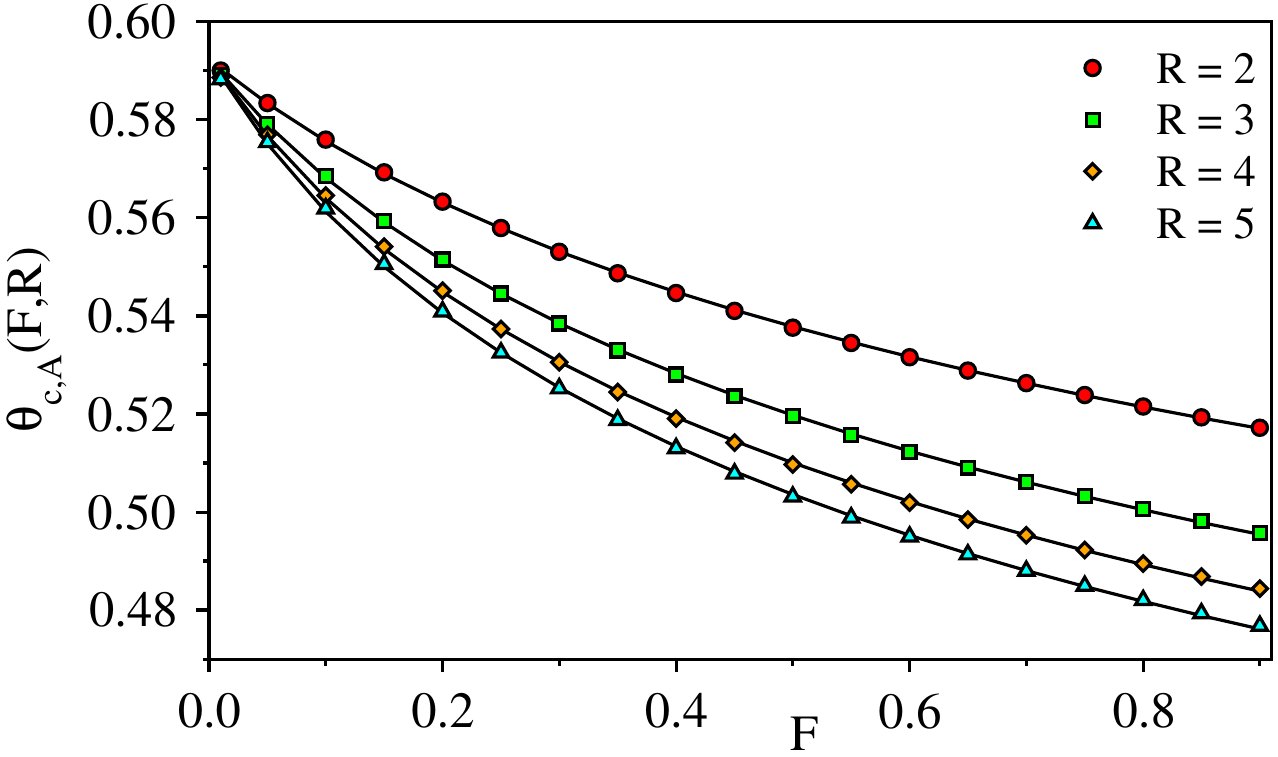}
\caption{Percolation threshold $\theta_{\mathrm{c},\mathrm{A}}(F,R)$ 
of A-particles as a function of the relative flux $F$ for $R=2$, $3$, $4$, and $5$. The
data points are for $L=1024$ and averaged over (at least) $2 \times 10^5$ independent samples. The black solid lines are fits using Eq.~\ref{eq:theta_vs_F}.}
\label{fig:crit_mix}
\end{figure}
\begin{figure}[t]
\centering
\includegraphics[width=\linewidth]{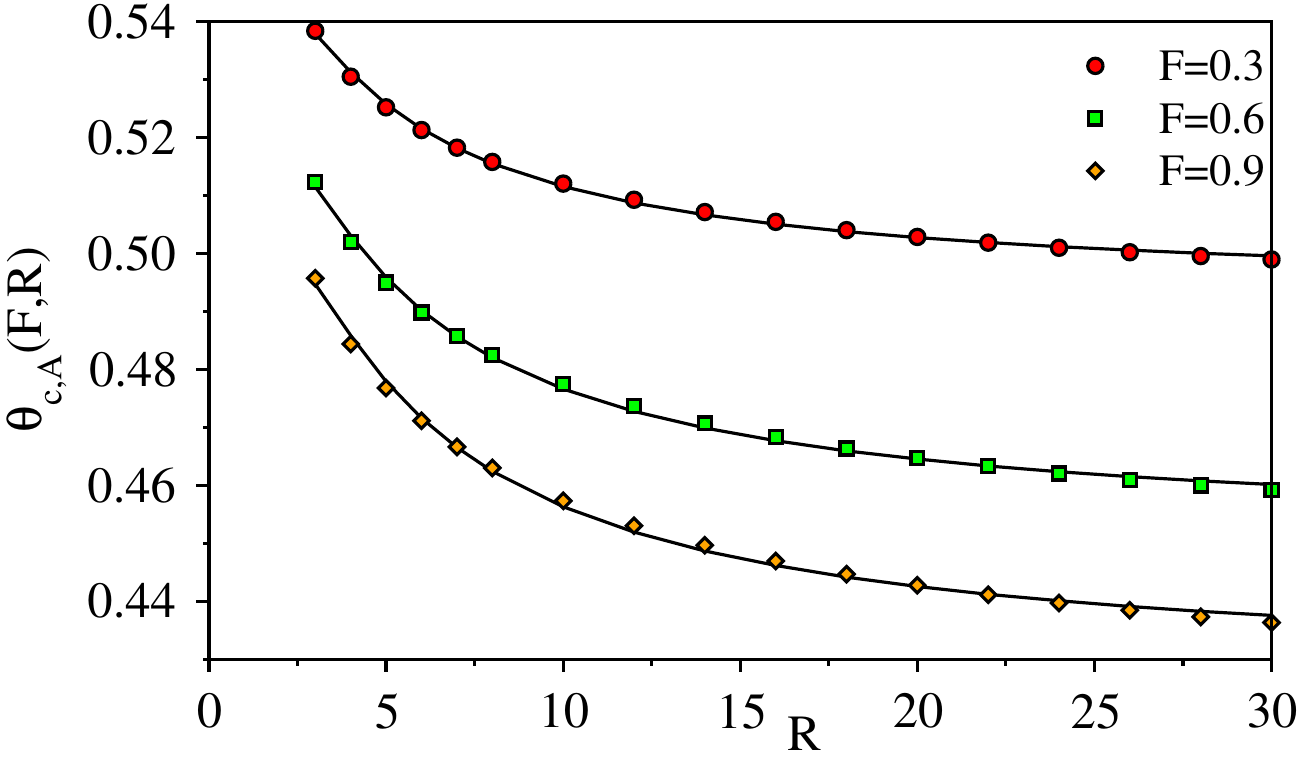}
\caption{Percolation threshold $\theta_{\mathrm{c},\mathrm{A}}(F,R)$ of 
A particles as a function of the aspect ratio $R$ for $F=0.3$, $0.6$, and
$0.9$. The values are obtained for $L=1024$ and averaged over (at least) $2 \times 10^5$ independent samples. The black solid lines represent the fit of the data using the functional form of} Eq.~\ref{eq:jamcov}.
\label{fig:crit_mix2}
\end{figure}

Numerically, the percolation threshold $\theta_{\mathrm{c},\mathrm{A}}(F,R)$ for a given value of $F$ and $R$ is determined in the following way. We start from an initially empty lattice and simulate the kinetics of adsorption until a spanning cluster of A particles emerges. We identify such a cluster using the Newman-Ziff algorithm~\cite{Newman2000,newmanziff}. In Fig.~\ref{fig:crit_mix} we plot $\theta_{\mathrm{c},\mathrm{A}}(F,R)$ as a function of the relative flux $F$ for four different values of $R$, namely, $2$, $3$, $4$, and $5$. For $R$=1, $\theta_{c,A}=p_c$ for all values of
$F<F_{\mathrm{c},\mathrm{A}}$, where $p_c$ is the percolation threshold on a square lattice. For $R>1$, it is observed that even the presence of a small 
fraction of B-particles favors the percolation of A-particles (lower percolation threshold) and, as $F$ increases, the percolation transition occurs even at a lower threshold coverage (Fig.~\ref{fig:crit_mix}). This effect becomes more pronounced the higher the value of $R$. The monotonic decay of the curve of $\theta_{\mathrm{c},\mathrm{A}}(F,R)$ with $F$
for a specific value of $R$ is described by the following functional form:
\begin{equation}
   \theta_{\mathrm{c},\mathrm{A}}(F,R) = \frac{d_1}{(d_2+F)^{d_3}}, 
\label{eq:theta_vs_F}   
\end{equation}
where $d_1$, $d_2$, and $d_3$ are fitting constants. The best fit of the data correspond to $d_1=0.5256(3)$, $0.5009(4)$, $0.4883(6)$, and $0.4801(8)$; 
$d_2=0.288(1)$, $0.208(1)$, $0.184(1)$, and $0.172(1)$; and $d_3=0.096(1)$, $0.107(1)$, $0.114(1)$, and 
$0.120(2)$ for $R=2$, $3$, $4$, and $5$, respectively.

Fig.~\ref{fig:crit_mix} also suggests that the distance between
two consecutive curves becomes smaller as $R$ is increased. This indicates that
for a infinitely large system, $\theta_{\mathrm{c},\mathrm{A}}(F,R)$ saturates to a $F$ dependent 
constant value as $R \to \infty$. To verify this, we plot 
$\theta_{\mathrm{c},\mathrm{A}}(F,R)$ as a function of $R$ for three different values of $F$ in Fig.~\ref{fig:crit_mix2}. Using the
same functional form as in Eq.~\ref{eq:jamcov}, we fit the data for all three values of $F$, which yields $\theta_{\mathrm{c},\mathrm{A}}(F,\infty)
= c_1 = 0.493(1)$, $0.451(1)$, and $0.427(2)$ for $F=0.3$, $0.6$, and $0.9$, respectively.

Finally, to verify the universality class of the percolation transition, we have estimated the correlation length exponent ($\nu$), the fractal dimension of the largest cluster ($d_f$), and the exponent ($d_l$) associated with the shortest path. For small $R$ these exponents are very much consistent with
the known values of the ordinary percolation exponents in two dimensions. For
large value of $R$ (e.g., $R=30$), though a slight deviation in the critical 
exponents values are observed numerically, they systematically converge with the lattice size towards the expected values for uncorrelated percolation. Therefore, we conclude that the model falls into the universality class of ordinary 
percolation for all model parameters.

\section{Conclusion}
\label{sec:conclusion}
We have studied the random sequential adsorption of a binary mixture of 
hard-core particles on the square lattice using Monte Carlo simulations. 
Adsorbing monomers and square-shaped particles with varying aspect ratio $R$ and 
relative flux $F$, the jamming and percolation properties 
have been investigated in the limit of irreversible adsorption. 

The jamming coverage $\theta_{\mathrm{J},\mathrm{B}}(F,R)$ of the larger particles has 
been calculated for different values of $R$ ranging from $2$ to $20$. We find that it decays with increasing $R$, which is well described by the relation given in 
Eq.~\ref{eq:jamcov} for a fixed $F$. The critical exponent 
$\nu_\mathrm{J}$ characterizing the jamming transition is found to be unitary in two dimensions. 

In the jammed state, the largest connected component of monomers 
undergoes a percolation transition for all finite values of $R$ for all values of $F<F_{\mathrm{c},\mathrm{A}}$. Above that value, for $R<4$, are the B-particles that percolate. A phase diagram is constructed on the $F$-$R$ plane (Fig.~\ref{fig:map}) showing clearly the regions of the percolating and 
non-percolating phases of monomers and squares in the mixture. 
During the kinetics of adsorption, the largest cluster 
of monomers in the mixture first percolates at a threshold value of 
its coverage $\theta_{\mathrm{c},\mathrm{A}}$, which is lower than the value 
$p_c$ of the site percolation threshold on the square lattice. Moreover, 
the percolation threshold $\theta_{\mathrm{c},\mathrm{A}}$ decreases monotonically as the size or the flux of the deposited squares is increased. Thus, in the presence of a second species, the percolation of monomers might be favored for a broad range of model parameters.

\emph{Acknowledgments.} The financial support from research grant of Universit\`{a} di Padova, Italy under the ``BIRD 2020 Orlandini'' fund is gratefully acknowledged. We also acknowledge financial support from the Portuguese Foundation for Science and Technology (FCT) under Contracts nos. PTDC/FIS-MAC/28146/2017 (LISBOA-01-0145-FEDER-028146), UIDB/00618/2020, and
UIDP/00618/2020.

\bibliography{bibli}

\begin{thebibliography}{57}%
\makeatletter
\providecommand \@ifxundefined [1]{%
 \@ifx{#1\undefined}
}%
\providecommand \@ifnum [1]{%
 \ifnum #1\expandafter \@firstoftwo
 \else \expandafter \@secondoftwo
 \fi
}%
\providecommand \@ifx [1]{%
 \ifx #1\expandafter \@firstoftwo
 \else \expandafter \@secondoftwo
 \fi
}%
\providecommand \natexlab [1]{#1}%
\providecommand \enquote  [1]{``#1''}%
\providecommand \bibnamefont  [1]{#1}%
\providecommand \bibfnamefont [1]{#1}%
\providecommand \citenamefont [1]{#1}%
\providecommand \href@noop [0]{\@secondoftwo}%
\providecommand \href [0]{\begingroup \@sanitize@url \@href}%
\providecommand \@href[1]{\@@startlink{#1}\@@href}%
\providecommand \@@href[1]{\endgroup#1\@@endlink}%
\providecommand \@sanitize@url [0]{\catcode `\\12\catcode `\$12\catcode
  `\&12\catcode `\#12\catcode `\^12\catcode `\_12\catcode `\%12\relax}%
\providecommand \@@startlink[1]{}%
\providecommand \@@endlink[0]{}%
\providecommand \url  [0]{\begingroup\@sanitize@url \@url }%
\providecommand \@url [1]{\endgroup\@href {#1}{\urlprefix }}%
\providecommand \urlprefix  [0]{URL }%
\providecommand \Eprint [0]{\href }%
\providecommand \doibase [0]{http://dx.doi.org/}%
\providecommand \selectlanguage [0]{\@gobble}%
\providecommand \bibinfo  [0]{\@secondoftwo}%
\providecommand \bibfield  [0]{\@secondoftwo}%
\providecommand \translation [1]{[#1]}%
\providecommand \BibitemOpen [0]{}%
\providecommand \bibitemStop [0]{}%
\providecommand \bibitemNoStop [0]{.\EOS\space}%
\providecommand \EOS [0]{\spacefactor3000\relax}%
\providecommand \BibitemShut  [1]{\csname bibitem#1\endcsname}%
\let\auto@bib@innerbib\@empty
\bibitem [{\citenamefont {Flory}(1939)}]{Flory1939}%
  \BibitemOpen
  \bibfield  {author} {\bibinfo {author} {\bibfnamefont {P.~J.}\ \bibnamefont
  {Flory}},\ }\bibfield  {booktitle} {\emph {\bibinfo {booktitle} {Journal of
  the American Chemical Society}},\ }\href {\doibase 10.1021/ja01875a053}
  {\bibfield  {journal} {\bibinfo  {journal} {Journal of the American Chemical
  Society}\ }\textbf {\bibinfo {volume} {61}},\ \bibinfo {pages} {1518}
  (\bibinfo {year} {1939})}\BibitemShut {NoStop}%
\bibitem [{\citenamefont {R\'enyi}(1958)}]{Renyi1958}%
  \BibitemOpen
  \bibfield  {author} {\bibinfo {author} {\bibfnamefont {A.}~\bibnamefont
  {R\'enyi}},\ }\href@noop {} {\bibfield  {journal} {\bibinfo  {journal} {Publ.
  Math. Inst. Hung. Acad. Sci}\ }\textbf {\bibinfo {volume} {3}},\ \bibinfo
  {pages} {109} (\bibinfo {year} {1958})}\BibitemShut {NoStop}%
\bibitem [{\citenamefont {Feder}(1980)}]{Feder1980}%
  \BibitemOpen
  \bibfield  {author} {\bibinfo {author} {\bibfnamefont {J.}~\bibnamefont
  {Feder}},\ }\href {\doibase https://doi.org/10.1016/0022-5193(80)90358-6}
  {\bibfield  {journal} {\bibinfo  {journal} {Journal of Theoretical Biology}\
  }\textbf {\bibinfo {volume} {87}},\ \bibinfo {pages} {237} (\bibinfo {year}
  {1980})}\BibitemShut {NoStop}%
\bibitem [{\citenamefont {Evans}(1993)}]{Evans1993}%
  \BibitemOpen
  \bibfield  {author} {\bibinfo {author} {\bibfnamefont {J.~W.}\ \bibnamefont
  {Evans}},\ }\href {\doibase 10.1103/RevModPhys.65.1281} {\bibfield  {journal}
  {\bibinfo  {journal} {Rev. Mod. Phys.}\ }\textbf {\bibinfo {volume} {65}},\
  \bibinfo {pages} {1281} (\bibinfo {year} {1993})}\BibitemShut {NoStop}%
\bibitem [{\citenamefont {Privman}(1994)}]{Privman1994}%
  \BibitemOpen
  \bibfield  {author} {\bibinfo {author} {\bibfnamefont {V.}~\bibnamefont
  {Privman}},\ }\href@noop {} {\bibfield  {journal} {\bibinfo  {journal}
  {Trends in Stat. Phys.}\ }\textbf {\bibinfo {volume} {1}},\ \bibinfo {pages}
  {89} (\bibinfo {year} {1994})}\BibitemShut {NoStop}%
\bibitem [{\citenamefont {Talbot}\ \emph {et~al.}(2000)\citenamefont {Talbot},
  \citenamefont {Tarjus}, \citenamefont {{Van Tassel}},\ and\ \citenamefont
  {Viot}}]{Talbot2000}%
  \BibitemOpen
  \bibfield  {author} {\bibinfo {author} {\bibfnamefont {J.}~\bibnamefont
  {Talbot}}, \bibinfo {author} {\bibfnamefont {G.}~\bibnamefont {Tarjus}},
  \bibinfo {author} {\bibfnamefont {P.}~\bibnamefont {{Van Tassel}}}, \ and\
  \bibinfo {author} {\bibfnamefont {P.}~\bibnamefont {Viot}},\ }\href {\doibase
  https://doi.org/10.1016/S0927-7757(99)00409-4} {\bibfield  {journal}
  {\bibinfo  {journal} {Colloids and Surfaces A: Physicochemical and
  Engineering Aspects}\ }\textbf {\bibinfo {volume} {165}},\ \bibinfo {pages}
  {287} (\bibinfo {year} {2000})}\BibitemShut {NoStop}%
\bibitem [{\citenamefont {Cadilhe}\ \emph {et~al.}(2007)\citenamefont
  {Cadilhe}, \citenamefont {Ara{\'{u}}jo},\ and\ \citenamefont
  {Privman}}]{Cadilhe2007}%
  \BibitemOpen
  \bibfield  {author} {\bibinfo {author} {\bibfnamefont {A.}~\bibnamefont
  {Cadilhe}}, \bibinfo {author} {\bibfnamefont {N.~A.~M.}\ \bibnamefont
  {Ara{\'{u}}jo}}, \ and\ \bibinfo {author} {\bibfnamefont {V.}~\bibnamefont
  {Privman}},\ }\href {\doibase 10.1088/0953-8984/19/6/065124} {\bibfield
  {journal} {\bibinfo  {journal} {J. Phys.: Condens. Matter}\ }\textbf
  {\bibinfo {volume} {19}},\ \bibinfo {pages} {065124} (\bibinfo {year}
  {2007})}\BibitemShut {NoStop}%
\bibitem [{\citenamefont {Torquato}\ and\ \citenamefont
  {Stillinger}(2010)}]{Torquato2010}%
  \BibitemOpen
  \bibfield  {author} {\bibinfo {author} {\bibfnamefont {S.}~\bibnamefont
  {Torquato}}\ and\ \bibinfo {author} {\bibfnamefont {F.~H.}\ \bibnamefont
  {Stillinger}},\ }\href {\doibase 10.1103/RevModPhys.82.2633} {\bibfield
  {journal} {\bibinfo  {journal} {Rev. Mod. Phys.}\ }\textbf {\bibinfo {volume}
  {82}},\ \bibinfo {pages} {2633} (\bibinfo {year} {2010})}\BibitemShut
  {NoStop}%
\bibitem [{\citenamefont {Kundu}\ \emph {et~al.}(2018)\citenamefont {Kundu},
  \citenamefont {Ara\'ujo},\ and\ \citenamefont {Manna}}]{Kundu2018}%
  \BibitemOpen
  \bibfield  {author} {\bibinfo {author} {\bibfnamefont {S.}~\bibnamefont
  {Kundu}}, \bibinfo {author} {\bibfnamefont {N.~A.~M.}\ \bibnamefont
  {Ara\'ujo}}, \ and\ \bibinfo {author} {\bibfnamefont {S.~S.}\ \bibnamefont
  {Manna}},\ }\href {\doibase 10.1103/PhysRevE.98.062118} {\bibfield  {journal}
  {\bibinfo  {journal} {Phys. Rev. E}\ }\textbf {\bibinfo {volume} {98}},\
  \bibinfo {pages} {062118} (\bibinfo {year} {2018})}\BibitemShut {NoStop}%
\bibitem [{\citenamefont {Furlan}\ \emph {et~al.}(2020)\citenamefont {Furlan},
  \citenamefont {dos Santos}, \citenamefont {Ziff},\ and\ \citenamefont
  {Dickman}}]{Furlan2020}%
  \BibitemOpen
  \bibfield  {author} {\bibinfo {author} {\bibfnamefont {A.~P.}\ \bibnamefont
  {Furlan}}, \bibinfo {author} {\bibfnamefont {D.~C.}\ \bibnamefont {dos
  Santos}}, \bibinfo {author} {\bibfnamefont {R.~M.}\ \bibnamefont {Ziff}}, \
  and\ \bibinfo {author} {\bibfnamefont {R.}~\bibnamefont {Dickman}},\ }\href
  {\doibase 10.1103/PhysRevResearch.2.043027} {\bibfield  {journal} {\bibinfo
  {journal} {Phys. Rev. Research}\ }\textbf {\bibinfo {volume} {2}},\ \bibinfo
  {pages} {043027} (\bibinfo {year} {2020})}\BibitemShut {NoStop}%
\bibitem [{\citenamefont {Kubiak}\ \emph {et~al.}(2016)\citenamefont {Kubiak},
  \citenamefont {Adamczyk},\ and\ \citenamefont {Cieśla}}]{Kubiak2016}%
  \BibitemOpen
  \bibfield  {author} {\bibinfo {author} {\bibfnamefont {K.}~\bibnamefont
  {Kubiak}}, \bibinfo {author} {\bibfnamefont {Z.}~\bibnamefont {Adamczyk}}, \
  and\ \bibinfo {author} {\bibfnamefont {M.}~\bibnamefont {Cieśla}},\ }\href
  {\doibase https://doi.org/10.1016/j.colsurfb.2015.11.052} {\bibfield
  {journal} {\bibinfo  {journal} {Colloids and Surfaces B: Biointerfaces}\
  }\textbf {\bibinfo {volume} {139}},\ \bibinfo {pages} {123} (\bibinfo {year}
  {2016})}\BibitemShut {NoStop}%
\bibitem [{\citenamefont {King}\ \emph {et~al.}(1974)\citenamefont {King},
  \citenamefont {Wells},\ and\ \citenamefont {Sheppard}}]{King1974}%
  \BibitemOpen
  \bibfield  {author} {\bibinfo {author} {\bibfnamefont {D.~A.}\ \bibnamefont
  {King}}, \bibinfo {author} {\bibfnamefont {M.~G.}\ \bibnamefont {Wells}}, \
  and\ \bibinfo {author} {\bibfnamefont {N.}~\bibnamefont {Sheppard}},\ }\href
  {\doibase 10.1098/rspa.1974.0120} {\bibfield  {journal} {\bibinfo  {journal}
  {Proceedings of the Royal Society of London. A. Mathematical and Physical
  Sciences}\ }\textbf {\bibinfo {volume} {339}},\ \bibinfo {pages} {245}
  (\bibinfo {year} {1974})}\BibitemShut {NoStop}%
\bibitem [{\citenamefont {Guo}\ \emph {et~al.}(1994)\citenamefont {Guo},
  \citenamefont {Bradley}, \citenamefont {Hopkinson},\ and\ \citenamefont
  {King}}]{Guo1994}%
  \BibitemOpen
  \bibfield  {author} {\bibinfo {author} {\bibfnamefont {X.-C.}\ \bibnamefont
  {Guo}}, \bibinfo {author} {\bibfnamefont {J.}~\bibnamefont {Bradley}},
  \bibinfo {author} {\bibfnamefont {A.}~\bibnamefont {Hopkinson}}, \ and\
  \bibinfo {author} {\bibfnamefont {D.}~\bibnamefont {King}},\ }\href {\doibase
  https://doi.org/10.1016/0039-6028(94)91382-X} {\bibfield  {journal} {\bibinfo
   {journal} {Surface Science}\ }\textbf {\bibinfo {volume} {310}},\ \bibinfo
  {pages} {163} (\bibinfo {year} {1994})}\BibitemShut {NoStop}%
\bibitem [{\citenamefont {Rodgers}\ and\ \citenamefont
  {Filipe}(1997)}]{Rodgers1997}%
  \BibitemOpen
  \bibfield  {author} {\bibinfo {author} {\bibfnamefont {G.~J.}\ \bibnamefont
  {Rodgers}}\ and\ \bibinfo {author} {\bibfnamefont {J.~A.~N.}\ \bibnamefont
  {Filipe}},\ }\href {\doibase 10.1088/0305-4470/30/10/021} {\bibfield
  {journal} {\bibinfo  {journal} {J. Phys. A: Math. Gen.}\ }\textbf {\bibinfo
  {volume} {30}},\ \bibinfo {pages} {3449} (\bibinfo {year}
  {1997})}\BibitemShut {NoStop}%
\bibitem [{\citenamefont {Finegold}\ and\ \citenamefont
  {Donnell}(1979)}]{Finegold1979}%
  \BibitemOpen
  \bibfield  {author} {\bibinfo {author} {\bibfnamefont {L.}~\bibnamefont
  {Finegold}}\ and\ \bibinfo {author} {\bibfnamefont {J.~T.}\ \bibnamefont
  {Donnell}},\ }\href {\doibase 10.1038/278443a0} {\bibfield  {journal}
  {\bibinfo  {journal} {Nature}\ }\textbf {\bibinfo {volume} {278}},\ \bibinfo
  {pages} {443} (\bibinfo {year} {1979})}\BibitemShut {NoStop}%
\bibitem [{\citenamefont {Feder}\ and\ \citenamefont
  {Giaever}(1980)}]{Feder1980-2}%
  \BibitemOpen
  \bibfield  {author} {\bibinfo {author} {\bibfnamefont {J.}~\bibnamefont
  {Feder}}\ and\ \bibinfo {author} {\bibfnamefont {I.}~\bibnamefont
  {Giaever}},\ }\href {\doibase https://doi.org/10.1016/0021-9797(80)90502-0}
  {\bibfield  {journal} {\bibinfo  {journal} {Journal of Colloid and Interface
  Science}\ }\textbf {\bibinfo {volume} {78}},\ \bibinfo {pages} {144}
  (\bibinfo {year} {1980})}\BibitemShut {NoStop}%
\bibitem [{\citenamefont {Onoda}\ and\ \citenamefont
  {Liniger}(1986)}]{Onoda1986}%
  \BibitemOpen
  \bibfield  {author} {\bibinfo {author} {\bibfnamefont {G.~Y.}\ \bibnamefont
  {Onoda}}\ and\ \bibinfo {author} {\bibfnamefont {E.~G.}\ \bibnamefont
  {Liniger}},\ }\href {\doibase 10.1103/PhysRevA.33.715} {\bibfield  {journal}
  {\bibinfo  {journal} {Phys. Rev. A}\ }\textbf {\bibinfo {volume} {33}},\
  \bibinfo {pages} {715} (\bibinfo {year} {1986})}\BibitemShut {NoStop}%
\bibitem [{\citenamefont {Joshi}\ \emph {et~al.}(2016)\citenamefont {Joshi},
  \citenamefont {Bargteil}, \citenamefont {Caciagli}, \citenamefont
  {Burelbach}, \citenamefont {Xing}, \citenamefont {Nunes}, \citenamefont
  {Pinto}, \citenamefont {Araújo}, \citenamefont {Brujic},\ and\ \citenamefont
  {Eiser}}]{Joshi2016}%
  \BibitemOpen
  \bibfield  {author} {\bibinfo {author} {\bibfnamefont {D.}~\bibnamefont
  {Joshi}}, \bibinfo {author} {\bibfnamefont {D.}~\bibnamefont {Bargteil}},
  \bibinfo {author} {\bibfnamefont {A.}~\bibnamefont {Caciagli}}, \bibinfo
  {author} {\bibfnamefont {J.}~\bibnamefont {Burelbach}}, \bibinfo {author}
  {\bibfnamefont {Z.}~\bibnamefont {Xing}}, \bibinfo {author} {\bibfnamefont
  {A.~S.}\ \bibnamefont {Nunes}}, \bibinfo {author} {\bibfnamefont {D.~E.~P.}\
  \bibnamefont {Pinto}}, \bibinfo {author} {\bibfnamefont {N.~A.~M.}\
  \bibnamefont {Araújo}}, \bibinfo {author} {\bibfnamefont {J.}~\bibnamefont
  {Brujic}}, \ and\ \bibinfo {author} {\bibfnamefont {E.}~\bibnamefont
  {Eiser}},\ }\href {\doibase 10.1126/sciadv.1600881} {\bibfield  {journal}
  {\bibinfo  {journal} {Science Advances}\ }\textbf {\bibinfo {volume} {2}},\
  \bibinfo {pages} {e1600881} (\bibinfo {year} {2016})}\BibitemShut {NoStop}%
\bibitem [{\citenamefont {Pinto}\ and\ \citenamefont
  {Ara\'ujo}(2018)}]{Pinto2018}%
  \BibitemOpen
  \bibfield  {author} {\bibinfo {author} {\bibfnamefont {D.~E.~P.}\
  \bibnamefont {Pinto}}\ and\ \bibinfo {author} {\bibfnamefont {N.~A.~M.}\
  \bibnamefont {Ara\'ujo}},\ }\href {\doibase 10.1103/PhysRevE.98.012125}
  {\bibfield  {journal} {\bibinfo  {journal} {Phys. Rev. E}\ }\textbf {\bibinfo
  {volume} {98}},\ \bibinfo {pages} {012125} (\bibinfo {year}
  {2018})}\BibitemShut {NoStop}%
\bibitem [{\citenamefont {Adamczyk}(2012)}]{Adamczyk2012}%
  \BibitemOpen
  \bibfield  {author} {\bibinfo {author} {\bibfnamefont {Z.}~\bibnamefont
  {Adamczyk}},\ }\href {\doibase https://doi.org/10.1016/j.cocis.2011.12.002}
  {\bibfield  {journal} {\bibinfo  {journal} {Current Opinion in Colloid \&
  Interface Science}\ }\textbf {\bibinfo {volume} {17}},\ \bibinfo {pages}
  {173} (\bibinfo {year} {2012})}\BibitemShut {NoStop}%
\bibitem [{\citenamefont {Krapivsky}(2020)}]{Krapivsky2020}%
  \BibitemOpen
  \bibfield  {author} {\bibinfo {author} {\bibfnamefont {P.~L.}\ \bibnamefont
  {Krapivsky}},\ }\href {\doibase 10.1103/PhysRevE.102.062108} {\bibfield
  {journal} {\bibinfo  {journal} {Phys. Rev. E}\ }\textbf {\bibinfo {volume}
  {102}},\ \bibinfo {pages} {062108} (\bibinfo {year} {2020})}\BibitemShut
  {NoStop}%
\bibitem [{\citenamefont {Subashiev}\ and\ \citenamefont
  {Luryi}(2007)}]{Subashiev2007}%
  \BibitemOpen
  \bibfield  {author} {\bibinfo {author} {\bibfnamefont {A.~V.}\ \bibnamefont
  {Subashiev}}\ and\ \bibinfo {author} {\bibfnamefont {S.}~\bibnamefont
  {Luryi}},\ }\href {\doibase 10.1103/PhysRevE.76.011128} {\bibfield  {journal}
  {\bibinfo  {journal} {Phys. Rev. E}\ }\textbf {\bibinfo {volume} {76}},\
  \bibinfo {pages} {011128} (\bibinfo {year} {2007})}\BibitemShut {NoStop}%
\bibitem [{\citenamefont {Talbot}\ and\ \citenamefont
  {Schaaf}(1989)}]{Talbot1989}%
  \BibitemOpen
  \bibfield  {author} {\bibinfo {author} {\bibfnamefont {J.}~\bibnamefont
  {Talbot}}\ and\ \bibinfo {author} {\bibfnamefont {P.}~\bibnamefont
  {Schaaf}},\ }\href {\doibase 10.1103/PhysRevA.40.422} {\bibfield  {journal}
  {\bibinfo  {journal} {Phys. Rev. A}\ }\textbf {\bibinfo {volume} {40}},\
  \bibinfo {pages} {422} (\bibinfo {year} {1989})}\BibitemShut {NoStop}%
\bibitem [{\citenamefont {Bartelt}\ and\ \citenamefont
  {Privman}(1991)}]{Bartelt1991}%
  \BibitemOpen
  \bibfield  {author} {\bibinfo {author} {\bibfnamefont {M.~C.}\ \bibnamefont
  {Bartelt}}\ and\ \bibinfo {author} {\bibfnamefont {V.}~\bibnamefont
  {Privman}},\ }\href {\doibase 10.1103/PhysRevA.44.R2227} {\bibfield
  {journal} {\bibinfo  {journal} {Phys. Rev. A}\ }\textbf {\bibinfo {volume}
  {44}},\ \bibinfo {pages} {R2227} (\bibinfo {year} {1991})}\BibitemShut
  {NoStop}%
\bibitem [{\citenamefont {Hassan}\ and\ \citenamefont
  {Kurths}(2001)}]{Hassan2001}%
  \BibitemOpen
  \bibfield  {author} {\bibinfo {author} {\bibfnamefont {M.~K.}\ \bibnamefont
  {Hassan}}\ and\ \bibinfo {author} {\bibfnamefont {J.}~\bibnamefont
  {Kurths}},\ }\href {\doibase 10.1088/0305-4470/34/37/307} {\bibfield
  {journal} {\bibinfo  {journal} {J. Phys. A: Math. Gen.}\ }\textbf {\bibinfo
  {volume} {34}},\ \bibinfo {pages} {7517} (\bibinfo {year}
  {2001})}\BibitemShut {NoStop}%
\bibitem [{\citenamefont {Hassan}\ \emph {et~al.}(2002)\citenamefont {Hassan},
  \citenamefont {Schmidt}, \citenamefont {Blasius},\ and\ \citenamefont
  {Kurths}}]{Hassan2002}%
  \BibitemOpen
  \bibfield  {author} {\bibinfo {author} {\bibfnamefont {M.~K.}\ \bibnamefont
  {Hassan}}, \bibinfo {author} {\bibfnamefont {J.}~\bibnamefont {Schmidt}},
  \bibinfo {author} {\bibfnamefont {B.}~\bibnamefont {Blasius}}, \ and\
  \bibinfo {author} {\bibfnamefont {J.}~\bibnamefont {Kurths}},\ }\href
  {\doibase 10.1103/PhysRevE.65.045103} {\bibfield  {journal} {\bibinfo
  {journal} {Phys. Rev. E}\ }\textbf {\bibinfo {volume} {65}},\ \bibinfo
  {pages} {045103} (\bibinfo {year} {2002})}\BibitemShut {NoStop}%
\bibitem [{\citenamefont {Doty}\ \emph {et~al.}(2002)\citenamefont {Doty},
  \citenamefont {Bonnecaze},\ and\ \citenamefont {Korgel}}]{Doty2002}%
  \BibitemOpen
  \bibfield  {author} {\bibinfo {author} {\bibfnamefont {R.~C.}\ \bibnamefont
  {Doty}}, \bibinfo {author} {\bibfnamefont {R.~T.}\ \bibnamefont {Bonnecaze}},
  \ and\ \bibinfo {author} {\bibfnamefont {B.~A.}\ \bibnamefont {Korgel}},\
  }\href {\doibase 10.1103/PhysRevE.65.061503} {\bibfield  {journal} {\bibinfo
  {journal} {Phys. Rev. E}\ }\textbf {\bibinfo {volume} {65}},\ \bibinfo
  {pages} {061503} (\bibinfo {year} {2002})}\BibitemShut {NoStop}%
\bibitem [{\citenamefont {Ara\'ujo}\ and\ \citenamefont
  {Cadilhe}(2006)}]{Araujo2006}%
  \BibitemOpen
  \bibfield  {author} {\bibinfo {author} {\bibfnamefont {N.~A.~M.}\
  \bibnamefont {Ara\'ujo}}\ and\ \bibinfo {author} {\bibfnamefont
  {A.}~\bibnamefont {Cadilhe}},\ }\href {\doibase 10.1103/PhysRevE.73.051602}
  {\bibfield  {journal} {\bibinfo  {journal} {Phys. Rev. E}\ }\textbf {\bibinfo
  {volume} {73}},\ \bibinfo {pages} {051602} (\bibinfo {year}
  {2006})}\BibitemShut {NoStop}%
\bibitem [{\citenamefont {Lon{\v c}arevi{\'c}}\ \emph
  {et~al.}(2007)\citenamefont {Lon{\v c}arevi{\'c}}, \citenamefont
  {Budinski-Petkovi{\'c}},\ and\ \citenamefont {Vrhovac}}]{Lon2007}%
  \BibitemOpen
  \bibfield  {author} {\bibinfo {author} {\bibfnamefont {I.}~\bibnamefont
  {Lon{\v c}arevi{\'c}}}, \bibinfo {author} {\bibfnamefont {L.}~\bibnamefont
  {Budinski-Petkovi{\'c}}}, \ and\ \bibinfo {author} {\bibfnamefont {S.~B.}\
  \bibnamefont {Vrhovac}},\ }\href@noop {} {\bibfield  {journal} {\bibinfo
  {journal} {The European Physical Journal E}\ }\textbf {\bibinfo {volume}
  {24}},\ \bibinfo {pages} {19} (\bibinfo {year} {2007})}\BibitemShut {NoStop}%
\bibitem [{\citenamefont {Dias}\ \emph {et~al.}(2012)\citenamefont {Dias},
  \citenamefont {Ara\'ujo},\ and\ \citenamefont {Cadilhe}}]{Dias2012}%
  \BibitemOpen
  \bibfield  {author} {\bibinfo {author} {\bibfnamefont {C.~S.}\ \bibnamefont
  {Dias}}, \bibinfo {author} {\bibfnamefont {N.~A.~M.}\ \bibnamefont
  {Ara\'ujo}}, \ and\ \bibinfo {author} {\bibfnamefont {A.}~\bibnamefont
  {Cadilhe}},\ }\href {\doibase 10.1103/PhysRevE.85.041120} {\bibfield
  {journal} {\bibinfo  {journal} {Phys. Rev. E}\ }\textbf {\bibinfo {volume}
  {85}},\ \bibinfo {pages} {041120} (\bibinfo {year} {2012})}\BibitemShut
  {NoStop}%
\bibitem [{\citenamefont {Dujak}\ \emph {et~al.}(2019)\citenamefont {Dujak},
  \citenamefont {Kara{\v{c}}}, \citenamefont {Budinski-Petkovi{\'{c}}},
  \citenamefont {Lon{\v{c}}arevi{\'{c}}}, \citenamefont {Jak{\v{s}}i{\'{c}}},\
  and\ \citenamefont {Vrhovac}}]{Dujak2019}%
  \BibitemOpen
  \bibfield  {author} {\bibinfo {author} {\bibfnamefont {D.}~\bibnamefont
  {Dujak}}, \bibinfo {author} {\bibfnamefont {A.}~\bibnamefont {Kara{\v{c}}}},
  \bibinfo {author} {\bibfnamefont {L.}~\bibnamefont
  {Budinski-Petkovi{\'{c}}}}, \bibinfo {author} {\bibfnamefont
  {I.}~\bibnamefont {Lon{\v{c}}arevi{\'{c}}}}, \bibinfo {author} {\bibfnamefont
  {Z.~M.}\ \bibnamefont {Jak{\v{s}}i{\'{c}}}}, \ and\ \bibinfo {author}
  {\bibfnamefont {S.~B.}\ \bibnamefont {Vrhovac}},\ }\href {\doibase
  10.1088/1742-5468/ab4588} {\bibfield  {journal} {\bibinfo  {journal} {J.
  Stat. Mech.}\ }\textbf {\bibinfo {volume} {2019}},\ \bibinfo {pages} {113210}
  (\bibinfo {year} {2019})}\BibitemShut {NoStop}%
\bibitem [{\citenamefont {Darjani}\ \emph {et~al.}(2021)\citenamefont
  {Darjani}, \citenamefont {Koplik}, \citenamefont {Pauchard},\ and\
  \citenamefont {Banerjee}}]{Darjani2021}%
  \BibitemOpen
  \bibfield  {author} {\bibinfo {author} {\bibfnamefont {S.}~\bibnamefont
  {Darjani}}, \bibinfo {author} {\bibfnamefont {J.}~\bibnamefont {Koplik}},
  \bibinfo {author} {\bibfnamefont {V.}~\bibnamefont {Pauchard}}, \ and\
  \bibinfo {author} {\bibfnamefont {S.}~\bibnamefont {Banerjee}},\ }\href
  {\doibase 10.1063/5.0039706} {\bibfield  {journal} {\bibinfo  {journal} {The
  Journal of Chemical Physics}\ }\textbf {\bibinfo {volume} {154}},\ \bibinfo
  {pages} {074705} (\bibinfo {year} {2021})}\BibitemShut {NoStop}%
\bibitem [{\citenamefont {Cornette}\ \emph {et~al.}(2006)\citenamefont
  {Cornette}, \citenamefont {Ramirez-Pastor},\ and\ \citenamefont
  {Nieto}}]{Cornette2006}%
  \BibitemOpen
  \bibfield  {author} {\bibinfo {author} {\bibfnamefont {V.}~\bibnamefont
  {Cornette}}, \bibinfo {author} {\bibfnamefont {A.}~\bibnamefont
  {Ramirez-Pastor}}, \ and\ \bibinfo {author} {\bibfnamefont {F.}~\bibnamefont
  {Nieto}},\ }\href {\doibase https://doi.org/10.1016/j.physleta.2006.01.007}
  {\bibfield  {journal} {\bibinfo  {journal} {Physics Letters A}\ }\textbf
  {\bibinfo {volume} {353}},\ \bibinfo {pages} {452 } (\bibinfo {year}
  {2006})}\BibitemShut {NoStop}%
\bibitem [{\citenamefont {Kondrat}(2006)}]{Kondrat2006}%
  \BibitemOpen
  \bibfield  {author} {\bibinfo {author} {\bibfnamefont {G.}~\bibnamefont
  {Kondrat}},\ }\href {\doibase 10.1063/1.2161206} {\bibfield  {journal}
  {\bibinfo  {journal} {The Journal of Chemical Physics}\ }\textbf {\bibinfo
  {volume} {124}},\ \bibinfo {pages} {054713} (\bibinfo {year}
  {2006})}\BibitemShut {NoStop}%
\bibitem [{\citenamefont {Centres}\ and\ \citenamefont
  {Ramirez-Pastor}(2015)}]{Centres2015}%
  \BibitemOpen
  \bibfield  {author} {\bibinfo {author} {\bibfnamefont {P.~M.}\ \bibnamefont
  {Centres}}\ and\ \bibinfo {author} {\bibfnamefont {A.~J.}\ \bibnamefont
  {Ramirez-Pastor}},\ }\href {\doibase 10.1088/1742-5468/2015/10/P10011}
  {\bibfield  {journal} {\bibinfo  {journal} {Journal of Statistical Mechanics:
  Theory and Experiment}\ }\textbf {\bibinfo {volume} {2015}},\ \bibinfo
  {pages} {P10011} (\bibinfo {year} {2015})}\BibitemShut {NoStop}%
\bibitem [{\citenamefont {Tarasevich}\ \emph {et~al.}(2015)\citenamefont
  {Tarasevich}, \citenamefont {Laptev}, \citenamefont {Vygornitskii},\ and\
  \citenamefont {Lebovka}}]{Tarasevich2015}%
  \BibitemOpen
  \bibfield  {author} {\bibinfo {author} {\bibfnamefont {Y.~Y.}\ \bibnamefont
  {Tarasevich}}, \bibinfo {author} {\bibfnamefont {V.~V.}\ \bibnamefont
  {Laptev}}, \bibinfo {author} {\bibfnamefont {N.~V.}\ \bibnamefont
  {Vygornitskii}}, \ and\ \bibinfo {author} {\bibfnamefont {N.~I.}\
  \bibnamefont {Lebovka}},\ }\href {\doibase 10.1103/PhysRevE.91.012109}
  {\bibfield  {journal} {\bibinfo  {journal} {Phys. Rev. E}\ }\textbf {\bibinfo
  {volume} {91}},\ \bibinfo {pages} {012109} (\bibinfo {year}
  {2015})}\BibitemShut {NoStop}%
\bibitem [{\citenamefont {Ramirez}\ \emph {et~al.}(2019)\citenamefont
  {Ramirez}, \citenamefont {Centres},\ and\ \citenamefont
  {Ramirez-Pastor}}]{Ramirez2019-2}%
  \BibitemOpen
  \bibfield  {author} {\bibinfo {author} {\bibfnamefont {L.~S.}\ \bibnamefont
  {Ramirez}}, \bibinfo {author} {\bibfnamefont {P.~M.}\ \bibnamefont
  {Centres}}, \ and\ \bibinfo {author} {\bibfnamefont {A.~J.}\ \bibnamefont
  {Ramirez-Pastor}},\ }\href {\doibase 10.1088/1742-5468/ab054d} {\bibfield
  {journal} {\bibinfo  {journal} {Journal of Statistical Mechanics: Theory and
  Experiment}\ }\textbf {\bibinfo {volume} {2019}},\ \bibinfo {pages} {033207}
  (\bibinfo {year} {2019})}\BibitemShut {NoStop}%
\bibitem [{\citenamefont {Palacios}\ and\ \citenamefont
  {Gomes}(2020)}]{Palacios2020}%
  \BibitemOpen
  \bibfield  {author} {\bibinfo {author} {\bibfnamefont {G.}~\bibnamefont
  {Palacios}}\ and\ \bibinfo {author} {\bibfnamefont {M.~A.~F.}\ \bibnamefont
  {Gomes}},\ }\href {\doibase 10.1088/1751-8121/ab9fb9} {\bibfield  {journal}
  {\bibinfo  {journal} {Journal of Physics A: Mathematical and Theoretical}\
  }\textbf {\bibinfo {volume} {53}},\ \bibinfo {pages} {375003} (\bibinfo
  {year} {2020})}\BibitemShut {NoStop}%
\bibitem [{\citenamefont {Kundu}\ and\ \citenamefont
  {Mandal}(2021)}]{Kundu2021}%
  \BibitemOpen
  \bibfield  {author} {\bibinfo {author} {\bibfnamefont {S.}~\bibnamefont
  {Kundu}}\ and\ \bibinfo {author} {\bibfnamefont {D.}~\bibnamefont {Mandal}},\
  }\href {\doibase 10.1103/PhysRevE.103.042134} {\bibfield  {journal} {\bibinfo
   {journal} {Phys. Rev. E}\ }\textbf {\bibinfo {volume} {103}},\ \bibinfo
  {pages} {042134} (\bibinfo {year} {2021})}\BibitemShut {NoStop}%
\bibitem [{\citenamefont {Tarjus}\ and\ \citenamefont
  {Talbot}(1991)}]{Tarjus1991}%
  \BibitemOpen
  \bibfield  {author} {\bibinfo {author} {\bibfnamefont {G.}~\bibnamefont
  {Tarjus}}\ and\ \bibinfo {author} {\bibfnamefont {J.}~\bibnamefont
  {Talbot}},\ }\href {\doibase 10.1088/0305-4470/24/16/006} {\bibfield
  {journal} {\bibinfo  {journal} {J. Phys. A: Math. Gen.}\ }\textbf {\bibinfo
  {volume} {24}},\ \bibinfo {pages} {L913} (\bibinfo {year}
  {1991})}\BibitemShut {NoStop}%
\bibitem [{\citenamefont {Budinski-Petkovi\ifmmode~\acute{c}\else \'{c}\fi{}}\
  \emph {et~al.}(2008)\citenamefont {Budinski-Petkovi\ifmmode~\acute{c}\else
  \'{c}\fi{}}, \citenamefont {Vrhovac},\ and\ \citenamefont {Lon\ifmmode
  \check{c}\else \v{c}\fi{}arevi\ifmmode~\acute{c}\else \'{c}\fi{}}}]{Lj2008}%
  \BibitemOpen
  \bibfield  {author} {\bibinfo {author} {\bibfnamefont {L.}~\bibnamefont
  {Budinski-Petkovi\ifmmode~\acute{c}\else \'{c}\fi{}}}, \bibinfo {author}
  {\bibfnamefont {S.~B.}\ \bibnamefont {Vrhovac}}, \ and\ \bibinfo {author}
  {\bibfnamefont {I.}~\bibnamefont {Lon\ifmmode \check{c}\else
  \v{c}\fi{}arevi\ifmmode~\acute{c}\else \'{c}\fi{}}},\ }\href {\doibase
  10.1103/PhysRevE.78.061603} {\bibfield  {journal} {\bibinfo  {journal} {Phys.
  Rev. E}\ }\textbf {\bibinfo {volume} {78}},\ \bibinfo {pages} {061603}
  (\bibinfo {year} {2008})}\BibitemShut {NoStop}%
\bibitem [{\citenamefont {Budinski-Petkovi\ifmmode~\acute{c}\else \'{c}\fi{}}\
  \emph {et~al.}(2011)\citenamefont {Budinski-Petkovi\ifmmode~\acute{c}\else
  \'{c}\fi{}}, \citenamefont {Lon\ifmmode \check{c}\else
  \v{c}\fi{}arevi\ifmmode~\acute{c}\else \'{c}\fi{}}, \citenamefont
  {Jak\ifmmode \check{s}\else \v{s}\fi{}i\ifmmode~\acute{c}\else \'{c}\fi{}},
  \citenamefont {Vrhovac},\ and\ \citenamefont {\ifmmode \check{S}\else
  \v{S}\fi{}vraki\ifmmode~\acute{c}\else \'{c}\fi{}}}]{Lj2011}%
  \BibitemOpen
  \bibfield  {author} {\bibinfo {author} {\bibfnamefont {L.}~\bibnamefont
  {Budinski-Petkovi\ifmmode~\acute{c}\else \'{c}\fi{}}}, \bibinfo {author}
  {\bibfnamefont {I.}~\bibnamefont {Lon\ifmmode \check{c}\else
  \v{c}\fi{}arevi\ifmmode~\acute{c}\else \'{c}\fi{}}}, \bibinfo {author}
  {\bibfnamefont {Z.~M.}\ \bibnamefont {Jak\ifmmode \check{s}\else
  \v{s}\fi{}i\ifmmode~\acute{c}\else \'{c}\fi{}}}, \bibinfo {author}
  {\bibfnamefont {S.~B.}\ \bibnamefont {Vrhovac}}, \ and\ \bibinfo {author}
  {\bibfnamefont {N.~M.}\ \bibnamefont {\ifmmode \check{S}\else
  \v{S}\fi{}vraki\ifmmode~\acute{c}\else \'{c}\fi{}}},\ }\href {\doibase
  10.1103/PhysRevE.84.051601} {\bibfield  {journal} {\bibinfo  {journal} {Phys.
  Rev. E}\ }\textbf {\bibinfo {volume} {84}},\ \bibinfo {pages} {051601}
  (\bibinfo {year} {2011})}\BibitemShut {NoStop}%
\bibitem [{\citenamefont {Meakin}\ and\ \citenamefont
  {Jullien}(1992)}]{Meakin1992}%
  \BibitemOpen
  \bibfield  {author} {\bibinfo {author} {\bibfnamefont {P.}~\bibnamefont
  {Meakin}}\ and\ \bibinfo {author} {\bibfnamefont {R.}~\bibnamefont
  {Jullien}},\ }\href {\doibase 10.1103/PhysRevA.46.2029} {\bibfield  {journal}
  {\bibinfo  {journal} {Phys. Rev. A}\ }\textbf {\bibinfo {volume} {46}},\
  \bibinfo {pages} {2029} (\bibinfo {year} {1992})}\BibitemShut {NoStop}%
\bibitem [{\citenamefont {Adamczyk}\ \emph {et~al.}(1997)\citenamefont
  {Adamczyk}, \citenamefont {Siwek}, \citenamefont {Zembala},\ and\
  \citenamefont {Weroński}}]{Adamczyk1997}%
  \BibitemOpen
  \bibfield  {author} {\bibinfo {author} {\bibfnamefont {Z.}~\bibnamefont
  {Adamczyk}}, \bibinfo {author} {\bibfnamefont {B.}~\bibnamefont {Siwek}},
  \bibinfo {author} {\bibfnamefont {M.}~\bibnamefont {Zembala}}, \ and\
  \bibinfo {author} {\bibfnamefont {P.}~\bibnamefont {Weroński}},\ }\href
  {\doibase https://doi.org/10.1006/jcis.1996.4540} {\bibfield  {journal}
  {\bibinfo  {journal} {Journal of Colloid and Interface Science}\ }\textbf
  {\bibinfo {volume} {185}},\ \bibinfo {pages} {236} (\bibinfo {year}
  {1997})}\BibitemShut {NoStop}%
\bibitem [{\citenamefont {Marques}\ \emph {et~al.}(2012)\citenamefont
  {Marques}, \citenamefont {Lima}, \citenamefont {Ara\'ujo},\ and\
  \citenamefont {Cadilhe}}]{Marques2012}%
  \BibitemOpen
  \bibfield  {author} {\bibinfo {author} {\bibfnamefont {J.~F.}\ \bibnamefont
  {Marques}}, \bibinfo {author} {\bibfnamefont {A.~B.}\ \bibnamefont {Lima}},
  \bibinfo {author} {\bibfnamefont {N.~A.~M.}\ \bibnamefont {Ara\'ujo}}, \ and\
  \bibinfo {author} {\bibfnamefont {A.}~\bibnamefont {Cadilhe}},\ }\href
  {\doibase 10.1103/PhysRevE.85.061122} {\bibfield  {journal} {\bibinfo
  {journal} {Phys. Rev. E}\ }\textbf {\bibinfo {volume} {85}},\ \bibinfo
  {pages} {061122} (\bibinfo {year} {2012})}\BibitemShut {NoStop}%
\bibitem [{\citenamefont {Brilliantov}\ \emph {et~al.}(1996)\citenamefont
  {Brilliantov}, \citenamefont {Andrienko}, \citenamefont {Krapivsky},\ and\
  \citenamefont {Kurths}}]{Brilliantov1996}%
  \BibitemOpen
  \bibfield  {author} {\bibinfo {author} {\bibfnamefont {N.~V.}\ \bibnamefont
  {Brilliantov}}, \bibinfo {author} {\bibfnamefont {Y.~A.}\ \bibnamefont
  {Andrienko}}, \bibinfo {author} {\bibfnamefont {P.~L.}\ \bibnamefont
  {Krapivsky}}, \ and\ \bibinfo {author} {\bibfnamefont {J.}~\bibnamefont
  {Kurths}},\ }\href {\doibase 10.1103/PhysRevLett.76.4058} {\bibfield
  {journal} {\bibinfo  {journal} {Phys. Rev. Lett.}\ }\textbf {\bibinfo
  {volume} {76}},\ \bibinfo {pages} {4058} (\bibinfo {year}
  {1996})}\BibitemShut {NoStop}%
\bibitem [{\citenamefont {Vieira}\ \emph {et~al.}(2011)\citenamefont {Vieira},
  \citenamefont {Gomes},\ and\ \citenamefont {{de Lima}}}]{Vieira2011}%
  \BibitemOpen
  \bibfield  {author} {\bibinfo {author} {\bibfnamefont {M.~C.}\ \bibnamefont
  {Vieira}}, \bibinfo {author} {\bibfnamefont {M.}~\bibnamefont {Gomes}}, \
  and\ \bibinfo {author} {\bibfnamefont {J.}~\bibnamefont {{de Lima}}},\ }\href
  {\doibase https://doi.org/10.1016/j.physa.2011.05.025} {\bibfield  {journal}
  {\bibinfo  {journal} {Physica A: Statistical Mechanics and its Applications}\
  }\textbf {\bibinfo {volume} {390}},\ \bibinfo {pages} {3404} (\bibinfo {year}
  {2011})}\BibitemShut {NoStop}%
\bibitem [{\citenamefont {Bonnier}\ \emph {et~al.}(1994)\citenamefont
  {Bonnier}, \citenamefont {Hontebeyrie}, \citenamefont {Leroyer},
  \citenamefont {Meyers},\ and\ \citenamefont {Pommiers}}]{Bonnier1994}%
  \BibitemOpen
  \bibfield  {author} {\bibinfo {author} {\bibfnamefont {B.}~\bibnamefont
  {Bonnier}}, \bibinfo {author} {\bibfnamefont {M.}~\bibnamefont
  {Hontebeyrie}}, \bibinfo {author} {\bibfnamefont {Y.}~\bibnamefont
  {Leroyer}}, \bibinfo {author} {\bibfnamefont {C.}~\bibnamefont {Meyers}}, \
  and\ \bibinfo {author} {\bibfnamefont {E.}~\bibnamefont {Pommiers}},\ }\href
  {\doibase 10.1103/PhysRevE.49.305} {\bibfield  {journal} {\bibinfo  {journal}
  {Phys. Rev. E}\ }\textbf {\bibinfo {volume} {49}},\ \bibinfo {pages} {305}
  (\bibinfo {year} {1994})}\BibitemShut {NoStop}%
\bibitem [{\citenamefont {Dickman}\ \emph {et~al.}(1991)\citenamefont
  {Dickman}, \citenamefont {Wang},\ and\ \citenamefont {Jensen}}]{Dickman1991}%
  \BibitemOpen
  \bibfield  {author} {\bibinfo {author} {\bibfnamefont {R.}~\bibnamefont
  {Dickman}}, \bibinfo {author} {\bibfnamefont {J.}~\bibnamefont {Wang}}, \
  and\ \bibinfo {author} {\bibfnamefont {I.}~\bibnamefont {Jensen}},\ }\href
  {\doibase 10.1063/1.460109} {\bibfield  {journal} {\bibinfo  {journal} {The
  Journal of Chemical Physics}\ }\textbf {\bibinfo {volume} {94}},\ \bibinfo
  {pages} {8252} (\bibinfo {year} {1991})}\BibitemShut {NoStop}%
\bibitem [{\citenamefont {Brosilow}\ \emph {et~al.}(1991)\citenamefont
  {Brosilow}, \citenamefont {Ziff},\ and\ \citenamefont
  {Vigil}}]{Brosilow1991}%
  \BibitemOpen
  \bibfield  {author} {\bibinfo {author} {\bibfnamefont {B.~J.}\ \bibnamefont
  {Brosilow}}, \bibinfo {author} {\bibfnamefont {R.~M.}\ \bibnamefont {Ziff}},
  \ and\ \bibinfo {author} {\bibfnamefont {R.~D.}\ \bibnamefont {Vigil}},\
  }\href {\doibase 10.1103/PhysRevA.43.631} {\bibfield  {journal} {\bibinfo
  {journal} {Phys. Rev. A}\ }\textbf {\bibinfo {volume} {43}},\ \bibinfo
  {pages} {631} (\bibinfo {year} {1991})}\BibitemShut {NoStop}%
\bibitem [{\citenamefont {Ramirez-Pastor}\ \emph {et~al.}(2019)\citenamefont
  {Ramirez-Pastor}, \citenamefont {Centres}, \citenamefont {Vogel},\ and\
  \citenamefont {Vald\'es}}]{Ramirez2019}%
  \BibitemOpen
  \bibfield  {author} {\bibinfo {author} {\bibfnamefont {A.~J.}\ \bibnamefont
  {Ramirez-Pastor}}, \bibinfo {author} {\bibfnamefont {P.~M.}\ \bibnamefont
  {Centres}}, \bibinfo {author} {\bibfnamefont {E.~E.}\ \bibnamefont {Vogel}},
  \ and\ \bibinfo {author} {\bibfnamefont {J.~F.}\ \bibnamefont {Vald\'es}},\
  }\href {\doibase 10.1103/PhysRevE.99.042131} {\bibfield  {journal} {\bibinfo
  {journal} {Phys. Rev. E}\ }\textbf {\bibinfo {volume} {99}},\ \bibinfo
  {pages} {042131} (\bibinfo {year} {2019})}\BibitemShut {NoStop}%
\bibitem [{\citenamefont {Pasinetti}\ \emph {et~al.}(2019)\citenamefont
  {Pasinetti}, \citenamefont {Ramirez}, \citenamefont {Centres}, \citenamefont
  {Ramirez-Pastor},\ and\ \citenamefont {Cwilich}}]{Pasinetti2019}%
  \BibitemOpen
  \bibfield  {author} {\bibinfo {author} {\bibfnamefont {P.~M.}\ \bibnamefont
  {Pasinetti}}, \bibinfo {author} {\bibfnamefont {L.~S.}\ \bibnamefont
  {Ramirez}}, \bibinfo {author} {\bibfnamefont {P.~M.}\ \bibnamefont
  {Centres}}, \bibinfo {author} {\bibfnamefont {A.~J.}\ \bibnamefont
  {Ramirez-Pastor}}, \ and\ \bibinfo {author} {\bibfnamefont {G.~A.}\
  \bibnamefont {Cwilich}},\ }\href {\doibase 10.1103/PhysRevE.100.052114}
  {\bibfield  {journal} {\bibinfo  {journal} {Phys. Rev. E}\ }\textbf {\bibinfo
  {volume} {100}},\ \bibinfo {pages} {052114} (\bibinfo {year}
  {2019})}\BibitemShut {NoStop}%
\bibitem [{\citenamefont {Nakamura}(1987)}]{Nakamura1987}%
  \BibitemOpen
  \bibfield  {author} {\bibinfo {author} {\bibfnamefont {M.}~\bibnamefont
  {Nakamura}},\ }\href {\doibase 10.1103/PhysRevA.36.2384} {\bibfield
  {journal} {\bibinfo  {journal} {Phys. Rev. A}\ }\textbf {\bibinfo {volume}
  {36}},\ \bibinfo {pages} {2384} (\bibinfo {year} {1987})}\BibitemShut
  {NoStop}%
\bibitem [{\citenamefont {Jacobsen}(2015)}]{Jacobsen2015}%
  \BibitemOpen
  \bibfield  {author} {\bibinfo {author} {\bibfnamefont {J.~L.}\ \bibnamefont
  {Jacobsen}},\ }\href {\doibase 10.1088/1751-8113/48/45/454003} {\bibfield
  {journal} {\bibinfo  {journal} {Journal of Physics A: Mathematical and
  Theoretical}\ }\textbf {\bibinfo {volume} {48}},\ \bibinfo {pages} {454003}
  (\bibinfo {year} {2015})}\BibitemShut {NoStop}%
\bibitem [{\citenamefont {Herrmann}\ \emph {et~al.}(1984)\citenamefont
  {Herrmann}, \citenamefont {Hong},\ and\ \citenamefont
  {Stanley}}]{Herrmann1984}%
  \BibitemOpen
  \bibfield  {author} {\bibinfo {author} {\bibfnamefont {H.~J.}\ \bibnamefont
  {Herrmann}}, \bibinfo {author} {\bibfnamefont {D.~C.}\ \bibnamefont {Hong}},
  \ and\ \bibinfo {author} {\bibfnamefont {H.~E.}\ \bibnamefont {Stanley}},\
  }\href {\doibase 10.1088/0305-4470/17/5/008} {\bibfield  {journal} {\bibinfo
  {journal} {Journal of Physics A: Mathematical and General}\ }\textbf
  {\bibinfo {volume} {17}},\ \bibinfo {pages} {L261} (\bibinfo {year}
  {1984})}\BibitemShut {NoStop}%
\bibitem [{\citenamefont {Newman}\ and\ \citenamefont
  {Ziff}(2000)}]{Newman2000}%
  \BibitemOpen
  \bibfield  {author} {\bibinfo {author} {\bibfnamefont {M.~E.~J.}\
  \bibnamefont {Newman}}\ and\ \bibinfo {author} {\bibfnamefont {R.~M.}\
  \bibnamefont {Ziff}},\ }\href {\doibase 10.1103/PhysRevLett.85.4104}
  {\bibfield  {journal} {\bibinfo  {journal} {Phys. Rev. Lett.}\ }\textbf
  {\bibinfo {volume} {85}},\ \bibinfo {pages} {4104} (\bibinfo {year}
  {2000})}\BibitemShut {NoStop}%
\bibitem [{\citenamefont {Newman}\ and\ \citenamefont
  {Ziff}(2001)}]{newmanziff}%
  \BibitemOpen
  \bibfield  {author} {\bibinfo {author} {\bibfnamefont {M.~E.~J.}\
  \bibnamefont {Newman}}\ and\ \bibinfo {author} {\bibfnamefont {R.~M.}\
  \bibnamefont {Ziff}},\ }\href {\doibase 10.1103/PhysRevE.64.016706}
  {\bibfield  {journal} {\bibinfo  {journal} {Phys. Rev. E}\ }\textbf {\bibinfo
  {volume} {64}},\ \bibinfo {pages} {016706} (\bibinfo {year}
  {2001})}\BibitemShut {NoStop}%
\end{thebibliography}%

\end{document}